\documentclass[aps,pra,twocolumn,a4paper,showpacs,10pt]{revtex4-2}
\usepackage{graphicx}
\usepackage{amsmath}
\usepackage{amssymb}
\usepackage{mathrsfs}
\usepackage{amsfonts}
\usepackage{dsfont}
\usepackage{dcolumn}
\usepackage{bm}
\usepackage{booktabs}
\usepackage{color}
\usepackage{xcolor}
\usepackage{colortbl}
\usepackage{mathtools}
\usepackage{upgreek}
\usepackage{cancel}

\newcommand{\di}{\mathrm{d}}
\newcommand{\bnabla}{\bm{\nabla}}
\providecommand{\abs}[1]{\left|#1\right|}

\begin{document}

\title{Helicity, chirality and spin of optical fields without vector potentials}

\author{Andrea Aiello}
\email{andrea.aiello@mpl.mpg.de} 
\affiliation{Max Planck Institute for the Science of Light, Staudtstrasse 2, 91058
Erlangen, Germany}


\date{\today}

\begin{abstract}
Helicity $H$, chirality $C$, and spin angular momentum $\mathbf{S}$ are three physical observables that play an important role in the study of optical fields.
These quantities are closely related, but their connection is hidden by the use of four different vector fields for their representation, namely, the electric and magnetic fields $\mathbf{E}$ and $\mathbf{B}$,  and the two transverse potential vectors $\mathbf{C}^\perp$ and $\mathbf{A}^\perp$. Helmholtz's decomposition theorem restricted to solenoidal vector fields, entails the introduction of a bona fide inverse curl operator, which permits one to express the above three quantities in terms of the observable electric and magnetic fields only. This yields clear expressions for $H, C$, and $\mathbf{S}$, which are automatically gauge-invariant and display electric-magnetic democracy.
\end{abstract}

\maketitle

\section{Introduction}\label{intro}

In both classical and quantum mechanics, physical systems can be characterized by conserved quantities, that is, observable quantities that do not change with time \cite{Coleman2019}. A conserved quantity $F = F[f]$ is a linear functional of a space- and time-dependent density $f=f(\mathbf{r},t)$ of the form
\begin{align}
F[f] = \int \di^3 \mathbf{r} \,  f(\mathbf{r},t) \, , \label{e10}
\end{align}
where $f(\mathbf{r},t)$ denotes either a scalar or a tensor function \cite{HatfieldBook}. The free electromagnetic field possesses an infinite set of conserved quantities associated with densities which are  bilinear functions of the field variables \cite{Jackson,Lipkin1964,Candlin1965,Kibble1965}.
Not all of these quantities have a clear physical meaning. Among the meaningful ones, the helicity $H=H[h]$, the chirality $C=C[\chi]$, and the spin angular momentum $\mathbf{S} = \mathbf{S}[\mathbf{s}]$ are particularly significant for the study of optical fields \cite{Barnett_2010,PhysRevA.83.021803,PhysRevA.85.063810,Cameron_2012,Cameron_2013,PhysRevA.86.042103,Barnett_2016,Galaverni2021,PhysRevA.105.023524}.
The corresponding densities $h=h(\mathbf{r},t), \chi=\chi(\mathbf{r},t)$, and $\mathbf{s} = \mathbf{s}(\mathbf{r},t)$ are given by
\begin{subequations}\label{e20}
\begin{align}
h & =  \frac{1}{2} \left[ \sqrt{\frac{\epsilon_0}{\mu_0}} \, \mathbf{A} \cdot \left( \bnabla \times \mathbf{A} \right) + \sqrt{\frac{\mu_0}{\epsilon_0}} \, \mathbf{C} \cdot \left( \bnabla \times \mathbf{C} \right)\right], \label{e20a} \\[6pt]
 \chi  & = \frac{\epsilon_0}{2} \Bigl[ \mathbf{E} \cdot \left( \bnabla \times \mathbf{E} \right) + c^2 \mathbf{B} \cdot \left( \bnabla \times \mathbf{B} \right) \Bigr] , \label{e20b} \\[6pt]
 \mathbf{s} & =  \frac{1}{2} \Bigl[  \epsilon_0 \, \mathbf{E} \times \mathbf{A} + \mathbf{B} \times \mathbf{C} \Bigr]  , \label{e20c}
\end{align}
\end{subequations}
where $\mathbf{E} = \mathbf{E}(\mathbf{r},t)$ and $\mathbf{B} = \mathbf{B}(\mathbf{r},t)$ are the electric and magnetic fields, respectively, and here and hereafter
\begin{subequations}\label{e25}
\begin{align}
\mathbf{C} & =  \mathbf{C}^\perp(\mathbf{r},t), \label{e25a} \\[6pt]
\mathbf{A}  & = \mathbf{A}^\perp(\mathbf{r},t), \label{e25b}
\end{align}
\end{subequations}
are the gauge-independent \textit{transverse} (or \textit{solenoidal}) parts of the vector potentials implicitly defined in the Coulomb gauge by
\begin{subequations}
\begin{align}
\mathbf{E} & = - \frac{1}{\epsilon_0} \bnabla \times \mathbf{C} \, , \label{e30a} \\[6pt]
\mathbf{B} & = \bnabla \times \mathbf{A}   \,  . \label{e30b}
\end{align}
\end{subequations}
We remark that from \eqref{e20} and \eqref{e25} it follows that $h, \chi$, and $\mathbf{s}$ are gauge-invariant quantities \cite{PhysRevA.86.013845,Barnett_2016}.

Substituting \eqref{e30a} and \eqref{e30b} into \eqref{e20a}-\eqref{e20c}, one could express $h, \chi$, and $\mathbf{s}$  in terms of the gauge fields $\mathbf{C}$ and $\mathbf{A}$ only.
However, as $\mathbf{C}$ and $\mathbf{A}$ are gauge-invariant but  not directly observable, it would be more appealing instead to write $h, \chi$, and $\mathbf{s}$ as functions of the physical fields $\mathbf{E}$ and $\mathbf{B}$ solely.
For this we would need to invert \eqref{e30a} and \eqref{e30b} to obtain,
\begin{subequations}
\begin{align}
\mathbf{C} & =  - \epsilon_0 \left( \bnabla \times \right)^{-1} \mathbf{E} \, , \label{e32a} \\[6pt]
\mathbf{A} & =  \left( \bnabla \times \right)^{-1} \mathbf{B}   \,  , \label{e32b}
\end{align}
\end{subequations}
where $(\bnabla \times)^{-1}$ would denote the formal inverse curl operator.
In this paper, we aim at determining such representation of helicity, chirality, and spin densities in terms of physical fields $\mathbf{E}$ and $\mathbf{B}$ only, to highlight their connection and to present a uniform view of these three fundamental quantities that are conserved in vacuum. Moreover, since the equations describing these quantities are expressed in terms of physical fields only, they could also be considered in order to extend the present study to non-vacuum cases, where charges and currents are present so that the dual symmetry of Maxwell's equations is broken and the electric vector potential cannot be introduced. However, such study would be beyond the scope of the present paper.

To achieve our goal, we invert \eqref{e30a} and \eqref{e30b} using Helmholtz's decomposition theorem, to obtain $\mathbf{C}=\mathbf{C}[\mathbf{E}]$ and $\mathbf{A}=\mathbf{A}[\mathbf{B}]$.
Inserting these functionals into \eqref{e20a}-\eqref{e20c}, we obtain the sought physical-field representation in real space. Then, we write $h, \chi$,and $\mathbf{s}$ in reciprocal space by means of the Fourier transforms of the physical fields. This provides further information on their connection and shows that, unlike the corresponding conserved quantities, these densities do not separate into the sum of right-handed and left-handed terms.
 {Finally, in the penultimate section we show that our findings for the gauge-invariant optical spin $\mathbf{S}$ are in agreement with previously established results \cite{Barnett_2016}.}
The paper is completed by two appendices of a mainly didactic nature, in which all the details of the calculations omitted in the main text are presented.

\section{Helmholtz's decomposition theorem for solenoidal vector fields}\label{Helmholtz}

In this section we briefly illustrate the Helmholtz decomposition theorem for solenoidal vector fields, following closely Sec. 6-3 in \cite{Klauder} and appendix F in \cite{kristensson2016scattering}.
For the sake of definiteness, we consider the magnetic field $\mathbf{B}=\mathbf{B}(\mathbf{r},t)$  and the transverse part of the potential vector $\mathbf{A}=\mathbf{A}^\perp (\mathbf{r},t)$ as prototypical solenoidal fields connected by the relation
\begin{align}
\mathbf{B}  = \bnabla  \times \mathbf{A}   \, . \label{e60}
\end{align}
We assume that both $\mathbf{A}$ and $\mathbf{B}$ go to zero faster than $1/r$ as $r \to \infty$ \cite{Zangwill}, where $r \equiv \abs{\mathbf{r}}$.

Let $\tilde{\mathbf{A}} = \tilde{\mathbf{A}}^\perp (\mathbf{k},t)$ and $\tilde{\mathbf{B}} = \tilde{\mathbf{B}}(\mathbf{k},t)$ be the spatial Fourier transform of $\mathbf{A} = \mathbf{A}^\perp (\mathbf{r},t)$ and $\mathbf{B}(\mathbf{r},t)$, respectively.
They are defined by
\begin{subequations}
\begin{align}
\mathbf{A}^\perp (\mathbf{r},t) & =   \int \frac{\di^3 \mathbf{k}}{\left( 2 \pi \right)^{3/2}} \, \tilde{\mathbf{A}}^\perp (\mathbf{k},t) \exp \left(  i  \mathbf{k} \cdot \mathbf{r} \right) , \label{e40a} \\[6pt]
\mathbf{B}(\mathbf{r},t) & =   \int \frac{\di^3 \mathbf{k}}{\left( 2 \pi \right)^{3/2}} \, \tilde{\mathbf{B}}(\mathbf{k},t) \exp \left(  i  \mathbf{k} \cdot \mathbf{r} \right). \label{e40b}
\end{align}
\end{subequations}
From $\bnabla \cdot \mathbf{A} = 0 = \bnabla \cdot \mathbf{B}$, it follows that
\begin{align}
\mathbf{k} \cdot \tilde{\mathbf{A}} =0=  \mathbf{k} \cdot
\tilde{\mathbf{B}} \, . \label{e50}
\end{align}
Our goal is to find $\mathbf{A}^\perp (\mathbf{r},t)$ given $\mathbf{B} (\mathbf{r},t)$, that is, to give a meaning to the symbolic equation
\begin{align}
\mathbf{A}^\perp (\mathbf{r},t)  = (\bnabla \times)^{-1} \mathbf{B} (\mathbf{r},t)   \, , \label{e70}
\end{align}
where $(\bnabla \times)^{-1}$ denotes the formal inverse curl operator.
Substituting \eqref{e40a} and \eqref{e40b} into \eqref{e60}, and using
\begin{align}
\bnabla  \times \left[ \bm{a} \exp \left( i \mathbf{k} \cdot \mathbf{r} \right)  \right]  = i \left( \mathbf{k} \times \bm{a} \right)\exp \left( i \mathbf{k} \cdot \mathbf{r} \right)\, , \label{e80}
\end{align}
where $\bm{a}$ is an arbitrary constant ($\mathbf{r}$-independent) vector, we obtain
\begin{align}
\tilde{\mathbf{B}} = & \; i \, \mathbf{k}  \times \tilde{\mathbf{A}} \nonumber \\[6pt]
= & \; i \left(
             \begin{array}{ccc}
               0 & -k_3 & k_2 \\
               k_3 & 0 & -k_1 \\
               -k_2 & k_1 & 0 \\
             \end{array}
        \right)
\left(
  \begin{array}{c}
  \tilde{A}_1 \\
  \tilde{A}_2 \\
  \tilde{A}_3 \\
  \end{array}
\right) \, .
  \label{e90}
\end{align}
The determinant of the antisymmetric matrix above is equal to zero, therefore we cannot obtain $\tilde{\mathbf{A}}$ directly from \eqref{e90} by matrix inversion. However, multiplying both sides of \eqref{e90} by $i \, \mathbf{k} \times $, and using $\bm{a}  \times \left( \bm{b}  \times \bm{c} \right) = \left( \bm{a} \cdot \bm{c} \right) \bm{b} - \left( \bm{a} \cdot \bm{b} \right) \bm{c}$, we find
\begin{align}
i \, \mathbf{k}  \times \tilde{\mathbf{B}} = & \; - \mathbf{k}  \times \left( \mathbf{k}  \times \tilde{\mathbf{A}} \right) \nonumber \\[6pt]
= & \; - \Bigl[ \underbrace{( \mathbf{k} \cdot \tilde{\mathbf{A}} )}_{= \, 0} \mathbf{k} - ( \mathbf{k} \cdot \mathbf{k}  ) \tilde{\mathbf{A}} \Bigr] \nonumber \\[6pt]
 = & \; k^2  \tilde{\mathbf{A}}, \label{e100}
\end{align}
where $k^2 = \mathbf{k} \cdot \mathbf{k} = |\mathbf{k}|^2$, and \eqref{e50} has been used. From \eqref{e100} it immediately follows that
\begin{align}
\tilde{\mathbf{A}} =  i \, \frac{\mathbf{k}  \times \tilde{\mathbf{B}}}{k^2} \, . \label{e110}
\end{align}
Substituting this equation into \eqref{e40a}, we obtain
\begin{align}
\mathbf{A}^\perp (\mathbf{r},t) & =   \int \frac{\di^3 \mathbf{k}}{\left( 2 \pi \right)^{3/2}} \, \left[ i \, \frac{\mathbf{k} \times \tilde{\mathbf{B}}(\mathbf{k},t)}{k^2} \right] \exp \left(  i   \,  \mathbf{k} \cdot \mathbf{r} \right) \nonumber \\[6pt]
& =   \bnabla  \times \int \frac{\di^3 \mathbf{k}}{\left( 2 \pi \right)^{3/2}} \,  \frac{1}{k^2} \,  \tilde{\mathbf{B}} (\mathbf{k},t)\, \exp \left(  i \, \mathbf{k} \cdot \mathbf{r} \right) \, ,  \label{e120}
\end{align}
where \eqref{e80} has been used.  {Next, we use Eq. (6) in \cite{PhysRevA.79.032112} (with the role of $\mathbf{k}$ and $\mathbf{r} - \mathbf{r}'$ exchanged),} to write $1/k^2$ as
\begin{align}
 \frac{1}{k^2 }  = & \; \int  \frac{\di^3 \mathbf{r}'}{4 \pi} \, \frac{\exp \left[- i   \, \mathbf{k} \cdot \left(\mathbf{r} - \mathbf{r}' \right)\right]}{\left|\mathbf{r} - \mathbf{r}' \right|} \, . \label{e130}
\end{align}
Then, using the definition \eqref{e40b}, we can rewrite \eqref{e120} as
\begin{align}
\mathbf{A}^\perp (\mathbf{r},t) = & \;   \bnabla  \times \int \frac{\di^3 \mathbf{k}}{\left( 2 \pi \right)^{3/2}} \,   \int \frac{ \di^3 \mathbf{r}'}{4 \pi} \, \frac{\exp \left(  i \,  \mathbf{k} \cdot  \mathbf{r}' \right)}{\left|\mathbf{r} - \mathbf{r}' \right|}\,  \tilde{\mathbf{B}} (\mathbf{k},t) \nonumber \\[6pt]
= & \;   \bnabla  \times \int  \frac{\di^3 \mathbf{r}'}{4 \pi} \, \frac{\mathbf{B}(\mathbf{r}',t)}{\left|\mathbf{r} - \mathbf{r}' \right|} \, . \label{e140}
\end{align}
This equation can be cast in a different but  equivalent form using the vector identity
\begin{align}
\bnabla  \times \left( \phi \, \bm{a} \right) = \phi \left( \bnabla  \times \bm{a}\right) - \bm{a}  \times \left( \bnabla \phi \right) \, , \label{e150}
\end{align}
and
\begin{align}
\bnabla f (\mathbf{r} - \mathbf{r}') = -\bnabla' f (\mathbf{r} - \mathbf{r}') \, , \label{e160}
\end{align}
where here and hereafter $\bnabla'$  denotes the gradient with respect to the primed coordinates $\mathbf{r}' = (x',y',z')$, with $f$  an arbitrary function. Applying \eqref{e150} and \eqref{e160} to the integrand in \eqref{e140}, we find
\begin{align}
\bnabla  \times \left[ \frac{\mathbf{B}(\mathbf{r}',t)}{\left|\mathbf{r} - \mathbf{r}' \right|} \right] = \frac{\bnabla'  \times \mathbf{B}(\mathbf{r}',t)}{\left|\mathbf{r} - \mathbf{r}' \right|}
- \bnabla'  \times \left[ \frac{\mathbf{B}(\mathbf{r}',t)}{\left|\mathbf{r} - \mathbf{r}' \right|} \right] \, . \label{e170}
\end{align}
Inserting \eqref{e170} in \eqref{e140} yields two terms:
\begin{align}
\mathbf{A}^\perp (\mathbf{r},t) = & \;   \int \frac{\di^3 \mathbf{r}'}{4 \pi} \, \frac{\bnabla'  \times \, \mathbf{B}(\mathbf{r}',t)}{\left|\mathbf{r} - \mathbf{r}'  \right|} \nonumber \\[6pt]
& - \underbrace{\int \frac{\di^3 \mathbf{r}'}{4 \pi} \bnabla'  \times \left[ \frac{\mathbf{B}(\mathbf{r}',t)}{\left|\mathbf{r} - \mathbf{r}' \right|} \right]}_{\mathrm{surface \; term} \; = \; 0} \, . \label{e180}
\end{align}
The last volume integral in \eqref{e180} can be written as a surface integral with the surface of integration lying at infinity, and it  vanishes if $\mathbf{B}(\mathbf{r},t)$ goes to zero faster than $1/r$ as $r \to \infty$.

Thus, from \eqref{e140} and \eqref{e180}, it follows that for any solenoidal field $\mathbf{G}=\mathbf{G}^\perp (\mathbf{r},t)$, we can write
\begin{align}
 (\bnabla \times)^{-1} \mathbf{G}^\perp  (\mathbf{r},t) = & \;   \bnabla  \times \int  \frac{\di^3 \mathbf{r}'}{4 \pi} \, \frac{\mathbf{G}^\perp (\mathbf{r}',t)}{\left|\mathbf{r} - \mathbf{r}' \right|} \label{e182} \\[6pt]
 = & \;  \int \frac{\di^3 \mathbf{r}'}{4 \pi} \, \frac{\bnabla'  \times \, \mathbf{G}^\perp (\mathbf{r}',t)}{\left|\mathbf{r} - \mathbf{r}'  \right|} \, . \label{e184}
\end{align}
 {It is worth noting that this result in the form \eqref{e184}, was already given in 1962 by Belinfante \cite{PhysRev.128.2832}.}

Equations \eqref{e182} and \eqref{e184} give a faithful representation of the inverse curl operator for solenoidal fields.
Indeed, it is straightforward to prove that $\bnabla  \times \left[ (\bnabla \times)^{-1} \mathbf{G}^\perp  (\mathbf{r},t)\right] = \mathbf{G}^\perp    (\mathbf{r},t)$ using either \eqref{e182} or \eqref{e184}. The simplest way to prove this is by using the first equation:
\begin{align}
 \bnabla  \times \left[ (\bnabla \times)^{-1} \mathbf{G}^\perp  (\mathbf{r},t)\right] = & \;  \bnabla \times \left[ \bnabla  \times \int  \frac{\di^3 \mathbf{r}'}{4 \pi} \, \frac{\mathbf{G}^\perp(\mathbf{r}',t)}{\left|\mathbf{r} - \mathbf{r}' \right|} \right] \nonumber \\[6pt]
 = & \;  \bnabla \underbrace{\left[ \bnabla \cdot \int  \frac{\di^3 \mathbf{r}'}{4 \pi} \, \frac{\mathbf{G}^\perp(\mathbf{r}',t)}{\left|\mathbf{r} - \mathbf{r}' \right|} \right] }_{= \; 0}  \nonumber \\[6pt]
& - \nabla^2 \int  \frac{\di^3 \mathbf{r}'}{4 \pi} \, \frac{\mathbf{G}^\perp(\mathbf{r}',t)}{\left|\mathbf{r} - \mathbf{r}' \right|}   . \label{e190}
\end{align}
The first integral in \eqref{e190} is zero because from \eqref{e160} and
\begin{align}
\bnabla \cdot \left( \phi \, \bm{a} \right) = \bm{a} \cdot \left( \bnabla \phi \right) + \phi \left( \bnabla \cdot \bm{a}\right)  \, , \label{s40}
\end{align}
it follows that
\begin{align}
 \bnabla \cdot \int  \frac{\di^3 \mathbf{r}'}{4 \pi} \, \frac{\mathbf{G}^\perp(\mathbf{r}',t)}{\left|\mathbf{r} - \mathbf{r}' \right|}  \equiv & \;  \int  \frac{\di^3 \mathbf{r}'}{4 \pi} \, \mathbf{G}^\perp(\mathbf{r}',t) \cdot \bnabla \frac{1}{\left|\mathbf{r} - \mathbf{r}' \right|}  \nonumber \\[6pt]
 = & \; - \int  \frac{\di^3 \mathbf{r}'}{4 \pi} \, \mathbf{G}^\perp(\mathbf{r}',t) \cdot \bnabla' \frac{1}{\left|\mathbf{r} - \mathbf{r}' \right|}  \nonumber \\[6pt]
 = & \; - \underbrace{\int  \frac{\di^3 \mathbf{r}'}{4 \pi} \bnabla' \cdot \left[   \frac{\mathbf{G}^\perp(\mathbf{r}',t)}{\left|\mathbf{r} - \mathbf{r}' \right|} \right]}_{ \mathrm{surface \; term} \; = \; 0} \nonumber \\[6pt]
 & + \int  \frac{\di^3 \mathbf{r}'}{4 \pi} \, \frac{1}{\left|\mathbf{r} - \mathbf{r}' \right|}   \underbrace{\bnabla' \cdot \mathbf{G}^\perp(\mathbf{r}',t)}_{= \; 0} \nonumber \\[6pt]
 = & \; 0 \,  . \label{S50}
\end{align}
Finally, using the relation
\begin{align}
\nabla^2 \frac{1}{\left| \mathbf{r} - \mathbf{r}' \right|} = - 4 \pi \, \delta \left( \mathbf{r} - \mathbf{r}' \right)  \,  , \label{e200}
\end{align}
we can rewrite the second integral in \eqref{e190}  as
\begin{align}
- \nabla^2 \int  \frac{\di^3 \mathbf{r}'}{4 \pi} \, \frac{\mathbf{G}^\perp(\mathbf{r}',t)}{\left|\mathbf{r} - \mathbf{r}' \right|}= & \;  \int  \di^3 \mathbf{r}' \, \mathbf{G}^\perp(\mathbf{r}',t) \delta \left( \mathbf{r} - \mathbf{r}' \right) \nonumber \\[6pt]
= & \; \mathbf{G}^\perp(\mathbf{r},t) \, . \label{e210}
\end{align}
This completes the proof.

\section{Helicity, chirality, and spin  in real space}\label{HCSreal}

In this section we calculate $H[h]$, $C[\chi]$, and $\mathbf{S}[\mathbf{s}]$ in real space via \eqref{e10}. We evaluate explicitly the densities $h=h(\mathbf{r},t), \chi=\chi(\mathbf{r},t)$, and $\mathbf{s} = \mathbf{s}(\mathbf{r},t)$, in terms of the electric and magnetic fields only.

\subsection{Helicity}\label{Hreal}

 From the definition \eqref{e20a}, it follows that
\begin{align}
H & = \int \di^3 \mathbf{r} \, \frac{1}{2} \left[ \sqrt{\frac{\epsilon_0}{\mu_0}} \, \mathbf{A} \cdot \left( \bnabla \times \mathbf{A} \right) + \sqrt{\frac{\mu_0}{\epsilon_0}} \, \mathbf{C} \cdot \left( \bnabla  \times \mathbf{C} \right)\right], \nonumber \\[6pt]
  & = \frac{\epsilon_0}{2 c} \int \di^3 \mathbf{r} \, \Bigl\{\mathbf{E} \cdot \left[(\bnabla \times)^{-1} \mathbf{E}  \right] + c^2 \mathbf{B} \cdot \left[(\bnabla \times)^{-1} \mathbf{B}  \right] \Bigr\} \, , \label{H10}
\end{align}
where \eqref{e32a} and \eqref{e32b}, have been used. Next, we use \eqref{e184} with $\mathbf{G} = \mathbf{E}$ and $\mathbf{G} = \mathbf{B}$, respectively, to write $(\bnabla \times)^{-1} \mathbf{E} $ and $(\bnabla \times)^{-1} \mathbf{B} $ explicitly in \eqref{H10}, thus obtaining
\begin{widetext}
\begin{align}
 H = & \; \int \di^3 \mathbf{r} \left\{ \frac{\epsilon_0}{8 \pi c} \,\int \di^3 \mathbf{r}'  \, \frac{\mathbf{E}(\mathbf{r},t) \cdot \Bigl[ \bnabla'  \times \mathbf{E}(\mathbf{r}',t) \Bigr] + c^2 \mathbf{B}(\mathbf{r},t) \cdot \Bigl[ \bnabla'  \times \mathbf{B} (\mathbf{r}',t) \Bigr]}{\left|\mathbf{r} - \mathbf{r}' \right|} \right\} \, . \label{H20}
\end{align}
\end{widetext}
An equivalent  expression for $H$ was already given in \cite{Bialynicki_Birula_2014}.

\subsection{Chirality}\label{Creal}

From \eqref{e20b} it follows that $C$ is automatically fulfilling electric-magnetic democracy. However, to highlight its connection with the helicity and the spin, we can recast the expression \eqref{e20b} in a form similar to \eqref{H20}, as follows. We start from the defining equation
\begin{align}
C  = & \; \int \di^3 \mathbf{r} \left\{ \frac{\epsilon_0}{2}\Bigl[ \mathbf{E} \cdot \left( \bnabla  \times \mathbf{E} \right) + c^2 \mathbf{B} \cdot \left( \bnabla  \times \mathbf{B} \right) \Bigr]\right\}  \, , \label{C5}
\end{align}
and we rewrite it  as a double spatial integral with the help of the Dirac delta function:
\begin{align}
C = & \;  \frac{\epsilon_0}{2}   \int  \di^3 \mathbf{r}  \int  \di^3 \mathbf{r}'
\delta \left( \mathbf{r} - \mathbf{r}'  \right)  \mathbf{E}(\mathbf{r},t) \cdot \Bigl[ \bnabla'  \times \mathbf{E}(\mathbf{r}',t) \Bigr] \nonumber \\[6pt]
&  + \frac{\epsilon_0 \, c^2 }{2}   \int  \di^3 \mathbf{r}  \int  \di^3 \mathbf{r}' \Biggl\{
\delta \left( \mathbf{r} - \mathbf{r}'  \right) \mathbf{B}(\mathbf{r},t) \nonumber \\[6pt]
& \phantom{+ \frac{\epsilon_0 \, c^2 }{2}   \int  \di^3 \mathbf{r}  \int  \di^3 \mathbf{r}' \Biggl\{} \cdot \Bigl[ \bnabla'  \times \mathbf{B} (\mathbf{r}',t) \Bigr] \Biggr\} \nonumber \\[6pt]
 \equiv & \; C_E + C_B \, . \label{C10}
\end{align}
Let us consider first the electric-field contribution  $C_E$ in \eqref{C10}. Using
\begin{align}
 \delta \left( \mathbf{r} - \mathbf{r}' \right) = -\frac{1}{4 \pi} \,\nabla^2  \frac{1}{\left| \mathbf{r} - \mathbf{r}' \right|}  \,  , \label{C15}
\end{align}
and swapping the order of integration, we can rewrite $C_E$ as
\begin{align}
C_E = & \; - \, \frac{1}{4 \pi} \int \di^3 \mathbf{r}' \Biggl\{ \Bigl[ \bnabla'  \times \mathbf{E}(\mathbf{r}',t) \Bigr] \nonumber \\[6pt]
& \cdot \int  \di^3 \mathbf{r} \,
\mathbf{E}(\mathbf{r},t) \left( \nabla^2 \frac{1}{\left| \mathbf{r} - \mathbf{r}'\right|}\right) \Biggr\} . \label{C20}
\end{align}
Next, we notice that
\begin{align}
\mathbf{E} \left( \nabla^2 \frac{1}{R}\right) = & \;
\frac{1}{R} \left( \nabla^2 \mathbf{E}  \right) \nonumber \\[6pt]
& + \underbrace{\sum_{i=1}^3{\frac{\partial }{ \partial x_i} } \left[  \left( {\frac{\partial }{ \partial x_i} \,} \frac{1}{R}\right)\mathbf{E} - \frac{1}{R} \left( {\frac{\partial }{ \partial x_i} \,} \mathbf{E} \right)  \right] }_{ \mathrm{surface \; term \; that \; goes \;} \, \to \, 0, \; \mathrm{when \; integrated } },
  \label{C30}
\end{align}
where $R \equiv \left| \mathbf{r} - \mathbf{r}'\right|$. Inserting \eqref{C30} into \eqref{C20}, we obtain
\begin{align}
C_E = & \;  - \, \frac{1}{4 \pi}  \int  \di^3 \mathbf{r}  \int   \di^3 \mathbf{r}' \, \frac{\displaystyle \nabla^2 \mathbf{E}(\mathbf{r},t) \cdot \Bigl[ \bnabla'  \times \mathbf{E}(\mathbf{r}',t)  \Bigr]}{\left| \mathbf{r} - \mathbf{r}' \right|  } \nonumber \\[6pt]
= & \;  - \, \frac{1}{4 \pi  c^2}  \int  \di^3 \mathbf{r}  \int   \di^3 \mathbf{r}' \, \frac{\displaystyle \frac{\partial^2}{\partial t^2 } \, \mathbf{E}(\mathbf{r},t) \cdot \Bigl[ \bnabla'  \times \mathbf{E}(\mathbf{r}',t)  \Bigr]}{\left| \mathbf{r} - \mathbf{r}' \right|  } \, , \label{C40}
\end{align}
where the wave equation
\begin{align}
 \nabla^2  \mathbf{E}(\mathbf{r},t) - \frac{1}{c^2} \frac{\partial^2}{\partial t^2 } \, \mathbf{E}(\mathbf{r},t) = 0 \, , \label{c50}
\end{align}
has been used. The same procedure can be followed to calculate $C_B$, so that eventually we can write
\begin{widetext}
\begin{align}
 C =   \int \di^3 \mathbf{r} \left\{ -\frac{\epsilon_0}{8 \pi   c^2}\int \di^3 \mathbf{r}'   \,
\frac{ \displaystyle \frac{\partial^2 \mathbf{E}(\mathbf{r},t)}{ \partial t^2 } \cdot \left[ \bnabla'  \times \mathbf{E}(\mathbf{r}',t) \right] + c^2 \frac{\partial^2 \mathbf{B}(\mathbf{r},t)}{ \partial t^2 } \cdot \left[ \bnabla'  \times \mathbf{B} (\mathbf{r}',t) \right]}{\left| \mathbf{r} - \mathbf{r}' \right|  } \right\} \, . \label{C70}
\end{align}
\end{widetext}

\subsection{Spin}\label{Sreal}

 From  \eqref{e20c} it follows that
\begin{align}
 \mathbf{S}  = & \;  \frac{\epsilon_0}{2} \int \di^3 \mathbf{r} \, \Bigl\{  \mathbf{E}  \times \left[(\bnabla \times)^{-1} \mathbf{B}  \right] \nonumber \\[6pt]
  & \phantom{\;  \frac{\epsilon_0}{2} \int \di^3 \mathbf{r} \, \Bigl\{} - \mathbf{B}  \times \left[(\bnabla \times)^{-1} \mathbf{E}  \right] \Bigr\} \,  , \label{S10}
\end{align}
where \eqref{e32a} and \eqref{e32b} have been used. To begin with,  let us consider the first term in the equation above. Our goal is to evaluate the functional
\begin{multline} \label{S20}
\int \di^3 \mathbf{r} \,   \mathbf{E}  \times \left[(\bnabla \times)^{-1} \mathbf{B}  \right] \\[6pt]
=  \int \di^3 \mathbf{r} \,   \mathbf{E}  \times \left[ \bnabla  \times \int  \frac{\di^3 \mathbf{r}'}{4 \pi} \, \frac{\mathbf{B}(\mathbf{r}',t)}{\left|\mathbf{r} - \mathbf{r}' \right|} \right]  \\[6pt]
  \equiv  \int \di^3 \mathbf{r} \,   \mathbf{E}  \times \left( \bnabla  \times \mathbf{F} \right) \,  ,
\end{multline}
where \eqref{e140} has been used, and we have defined $\mathbf{F} = \mathbf{F}(\mathbf{r},t)$, as
\begin{align}
\mathbf{F}(\mathbf{r},t)= & \; \int  \frac{\di^3 \mathbf{r}'}{4 \pi} \, \frac{\mathbf{B}(\mathbf{r}',t)}{\left|\mathbf{r} - \mathbf{r}' \right|}  \,  . \label{S30}
\end{align}
Note that  \eqref{S50}  implies
\begin{align}
\bnabla \cdot \mathbf{F} = 0  \,  . \label{S32}
\end{align}

Next, we  show that
\begin{align}
\int \di^3 \mathbf{r} \,   \mathbf{E}  \times \left( \bnabla  \times \mathbf{F} \right)  = & \;  \int \di^3 \mathbf{r} \,  \left( \bnabla  \times \mathbf{E} \right)  \times \mathbf{F} \,  . \label{S60}
\end{align}
For this, first we notice that
\begin{align}
\bnabla \left( \mathbf{E} \cdot \mathbf{F} \right) = & \;   \mathbf{E}  \times \left( \bnabla  \times \mathbf{F} \right) +   \mathbf{F}  \times \left( \bnabla  \times \mathbf{E} \right) \nonumber \\[6pt]
& + \left( \mathbf{F} \cdot \bnabla \right) \mathbf{E} + \left( \mathbf{E} \cdot \bnabla \right) \mathbf{F} \,  . \label{S70}
\end{align}
The integral of $\bnabla \left( \mathbf{E} \cdot \mathbf{F} \right)$ with respect to $\di^3 \mathbf{r}$ is a surface term that goes to zero when the surface of integration goes to infinity. The integral of $ \left( \mathbf{F} \cdot \bnabla \right) \mathbf{E} + \left( \mathbf{E} \cdot \bnabla \right) \mathbf{F}$ is also  zero because
\begin{align}
\int  \di^3 \mathbf{r} \left( \mathbf{F} \cdot \bnabla \right) \mathbf{E}   = & \;  \int  \di^3 \mathbf{r}\,  F_i\left( \partial_i \mathbf{E} \right)
  \nonumber \\[6pt]
 = & \; \underbrace{\int  \di^3 \mathbf{r} \, \partial_i \left( F_i \mathbf{E} \right)}_{ \, \mathrm{surface \; term}\, = \, 0} - \int  \di^3 \mathbf{r}   \underbrace{\left( \bnabla \cdot \mathbf{F} \right)}_{= \, 0 \; \mathrm{from} \; \eqref{S32}} \! \! \! \mathbf{E} \nonumber \\[6pt]
 = & \; 0 \,  , \label{S80}
\end{align}
where $\partial_i = \partial / \partial x_i$, with $i = 1,2,3$, and summation over repeated indices is understood.
In the same way we can show that
\begin{align}
\int  \di^3 \mathbf{r} \left( \mathbf{E} \cdot \bnabla \right) \mathbf{F}   = & \;  0 \,  . \label{S90}
\end{align}
Thus, \eqref{S60} is demonstrated.

Now, using \eqref{S60} and Faraday's law
\begin{align}
 \bm{\nabla}  \times \mathbf{E}(\mathbf{r},t)  = & \;  -\frac{\partial \mathbf{B}(\mathbf{r},t)}{\partial t } \, , \label{S100}
\end{align}
we can rewrite \eqref{S20} as
\begin{multline}\label{S110}
\int \di^3 \mathbf{r} \,  \left( \bnabla  \times \mathbf{E} \right)  \times \mathbf{F} \\[6pt]
 =  \int \di^3 \mathbf{r} \, \left[  -\frac{\partial \mathbf{B}(\mathbf{r},t)}{\partial t } \right]   \times \int  \frac{\di^3 \mathbf{r}'}{4 \pi} \, \frac{\mathbf{B}(\mathbf{r}',t)}{\left|\mathbf{r} - \mathbf{r}' \right|}   \\[6pt]
 =   \frac{1}{4 \pi} \, \int \di^3 \mathbf{r} \int \di^3 \mathbf{r}'  \, \frac{\displaystyle \mathbf{B}(\mathbf{r}',t) \times \frac{\partial \mathbf{B}(\mathbf{r},t)}{\partial t }}{\left|\mathbf{r} - \mathbf{r}' \right|} .
\end{multline}
Following the same procedure as above, and using Amp\`{e}re's  law
\begin{align}
 \bm{\nabla}  \times \mathbf{B}(\mathbf{r},t)  = & \;  \frac{1}{c^2} \, \frac{\partial}{\partial t }\mathbf{E}(\mathbf{r},t) \, , \label{S120}
\end{align}
we can directly prove that
\begin{multline}\label{S130}
  \int \di^3 \mathbf{r} \,  \mathbf{B}  \times \left[(\bnabla \times)^{-1} \mathbf{E}  \right] \\[6pt]
 =  - \frac{1}{4 \pi \, c^2}  \int \di^3 \mathbf{r} \int \di^3 \mathbf{r}'  \, \frac{\displaystyle \mathbf{E}(\mathbf{r}',t)  \times \frac{\partial \mathbf{E}(\mathbf{r},t)}{\partial t }}{\left|\mathbf{r} - \mathbf{r}' \right|} \, .
\end{multline}
Finally, gathering \eqref{S10}, \eqref{S120}, and \eqref{S130}, we can write
\begin{widetext}
\begin{align}
 \mathbf{S} = & \;   \int \di^3 \mathbf{r} \left\{ \frac{\epsilon_0}{8 \pi  c^2} \int \di^3 \mathbf{r}'  \, \frac{\displaystyle \mathbf{E}(\mathbf{r},t) \times  \frac{\partial \mathbf{E}(\mathbf{r}',t)}{\partial t }
+ c^2 \,  \mathbf{B}(\mathbf{r},t)  \times \frac{\partial \mathbf{B}(\mathbf{r}',t)}{\partial t }
}{\left|\mathbf{r} - \mathbf{r}' \right|} \right\} \, . \label{S140}
\end{align}
\end{widetext}
By definition, this expression displays electric-magnetic democracy.

\subsection{Discussion}\label{Comp}

Equations \eqref{H20}, \eqref{C70}, and \eqref{S140} show that $h(\mathbf{r},t)$, $\chi(\mathbf{r},t)$, and $\mathbf{s}(\mathbf{r},t)$ all have  the same form, which is
\begin{align}\label{k20}
f(\mathbf{r},t) = & \;  \frac{\epsilon_0}{8 \pi c} \, \int \di^3 \mathbf{r}'  \, \frac{g_E(\mathbf{r}, \mathbf{r}',t)}{\left|\mathbf{r} - \mathbf{r}' \right|}  \nonumber \\[6pt]
& + \frac{\epsilon_0}{8 \pi c} \, \int \di^3 \mathbf{r}'  \, \frac{g_B(\mathbf{r}, \mathbf{r}',t)}{\left|\mathbf{r} - \mathbf{r}' \right|} \, ,
\end{align}
where $g \in \{ h, \chi, \mathbf{s}\}$, and
\begin{subequations}
\begin{align}
h_E(\mathbf{r},\mathbf{r}',t) = & \;   \mathbf{E}(\mathbf{r},t) \cdot \left[ \bnabla'  \times \mathbf{E}(\mathbf{r}',t) \right]  \, , \label{k10a} \\[6pt]
 \chi_E(\mathbf{r},\mathbf{r}',t)  = & \; - \frac{1}{c}
 \frac{\partial^2 \mathbf{E}(\mathbf{r},t)}{ \partial t^2 } \cdot \left[ \bnabla'  \times \mathbf{E}(\mathbf{r}',t) \right]  \, , \label{k10b} \\[6pt]
 \mathbf{s}_E(\mathbf{r},\mathbf{r}',t) = & \;  \frac{1}{c} \, \mathbf{E}(\mathbf{r},t) \times \frac{\partial \mathbf{E}(\mathbf{r}',t)}{\partial t } \,  . \label{k10c}
\end{align}
\end{subequations}
The corresponding magnetic densities $h_B, \chi_B$, and $\mathbf{s}_B$ can be obtained from \eqref{k10a}-\eqref{k10c} by replacing $\mathbf{E}$ with $c \mathbf{B}$ everywhere. Equations \eqref{k20} and \eqref{k10a}-\eqref{k10c} are the main result of this paper.

Next, we will make some remarks on the advantages of expressing $h, \chi$, and $\mathbf{s}$ in terms of physically observable electric and magnetic fields, compared to the traditional formulas \eqref{e20a}-\eqref{e20c}.
\begin{enumerate}
  \item In \eqref{k20}, electric-magnetic democracy is clearly displayed.
  \item From \eqref{k20},  the nonlocal nature of the three densities $h(\mathbf{r},t)$, $\chi(\mathbf{r},t)$, and $\mathbf{s}(\mathbf{r},t)$ is evident. The importance of this point has been thoroughly discussed by Bialynicki-Birula \cite{Bialynicki_Birula_2014}.
  \item By definition, equations \eqref{k20} and \eqref{k10a}-\eqref{k10c}, are manifestly gauge invariant.
  \item Comparing \eqref{k10a} and \eqref{k10b}, we can see directly why for monochromatic fields of frequency $\omega_0$, helicity and chirality are proportional to each other. In this case ${\partial^2 \mathbf{E}(\mathbf{r},t)}/{ \partial t^2 } = -\omega_0^2 \mathbf{E}(\mathbf{r},t)$, so that $ \chi_E(\mathbf{r},\mathbf{r}',t) =  (\omega_0^2/c) h_E(\mathbf{r},\mathbf{r}',t)$.
  \item The densities \eqref{k10a}-\eqref{k10c} are \emph{not} uniquely defined.  To obtain these equations, we repeatedly used integration by parts with respect to $\di^3 \mathbf{r}$, which caused the elimination of some surface terms. This is particularly evident for the chirality, where we have two distinct expressions for $\chi(\mathbf{r},t)$, given by \eqref{C5} and \eqref{C70}. Having different expressions for the densities $h(\mathbf{r},t)$, $\chi(\mathbf{r},t)$, and $\mathbf{s}(\mathbf{r},t)$, all giving the \emph{same} measurable conserved quantities $H,C$, and $\mathbf{S}$, respectively, allows for different views, all physically valid, of the same quantities. This produces deeper physical insights, as it will be particular evident in the next section where the densities are expressed in reciprocal (Fourier) space. A detailed discussion of why it is convenient  having non-uniquely defined densities can be found in sec. 5.4, p. 87 of \cite{Coleman2019}.
\end{enumerate}

\section{Helicity, chirality and spin of optical fields in reciprocal space}\label{HCSreciprocal}

Further insights into  $H[h]$, $C[\chi]$, and $\mathbf{S}[\mathbf{s}]$, come from their expressions in reciprocal space, obtained by substituting the Fourier transforms of $\mathbf{E}$ and $\mathbf{B}$ into Eqs. \eqref{k20} and \eqref{k10a}-\eqref{k10c}. As shown in Appendix \ref{appa}, Eqs. \eqref{a110}, \eqref{b50}, \eqref{b90},  and \eqref{s100} demonstrate that  it is possible to write
\begin{widetext}
\begin{align}\label{k30}
 f(\mathbf{r},t) =   -\frac{\epsilon_0}{c} \, \sum_{\sigma, \sigma' = \pm 1} \int \frac{\di^3  \mathbf{k}}{(2 \pi)^{3/2}}  \int \frac{\di^3 \mathbf{k}'}{(2 \pi)^{3/2}}  \, \frac{ e^{i   \mathbf{r} \cdot \left(\mathbf{k} - \mathbf{k}' \right)}}{\left| \mathbf{k}'\right|}   \, g_{\sigma \sigma '}(\mathbf{k},\mathbf{k}') A_{\sigma \sigma '}(\mathbf{k},\mathbf{k}',t) +  \mathrm{c.c.}  \, ,
\end{align}
where, as before, $g \in \{ h, \chi, \mathbf{s}\}$, and the common time-dependent term $A_{\sigma \sigma '}(\mathbf{k},\mathbf{k}',t) $ is defined by
\begin{align}\label{k50}
A_{\sigma \sigma '}(\mathbf{k},\mathbf{k}',t)
= & \;    a_\sigma(\mathbf{k})  a_{\sigma'}(-\mathbf{k}')\left(\frac{\sigma' - \sigma }{2}\right)   \exp\left[{- i \left( \omega + \omega' \right)t }\right] \nonumber \\[6pt]
& + a_\sigma(\mathbf{k})  a_{\sigma'}^*(\mathbf{k}') \left(\frac{\sigma' + \sigma}{2} \right) \exp\left[{- i \left( \omega - \omega' \right)t }\right]     \, ,
\end{align}
\end{widetext}
with $\omega = c \left| \mathbf{k} \right|$, $\omega' = c \left| \mathbf{k}' \right|$, and
\begin{subequations}
\begin{align}
h_{\sigma \sigma '}(\mathbf{k},\mathbf{k}') = & \;   \hat{\bm{\epsilon}}_{\sigma}(\hat{\mathbf{k}}) \cdot \hat{\bm{\epsilon}}_{\sigma'}^*(\hat{\mathbf{k}}') \; , \; \phantom{xxxxi} \mathrm{from} \; \eqref{a110}  , \label{k40a} \\[6pt]
\chi_{\sigma \sigma '}(\mathbf{k},\mathbf{k}')  = & \; c  \left| \mathbf{k} \right|^2 \hat{\bm{\epsilon}}_{\sigma}(\hat{\mathbf{k}}) \cdot \hat{\bm{\epsilon}}_{\sigma'}^*(\hat{\mathbf{k}}')\, , \;   \, \phantom{.} \mathrm{from} \; \eqref{b50}   , \label{k40b} \\[6pt]
\chi_{\sigma \sigma '}(\mathbf{k},\mathbf{k}') = & \; c  \left| \mathbf{k}' \right|^2 \hat{\bm{\epsilon}}_{\sigma}(\hat{\mathbf{k}}) \cdot \hat{\bm{\epsilon}}_{\sigma'}^*(\hat{\mathbf{k}}')\, , \; \,  \mathrm{from} \; \eqref{b90}   , \label{k40c} \\[6pt]
\mathbf{s}_{\sigma \sigma '}(\mathbf{k},\mathbf{k}') = & \; - i \sigma \, \hat{\bm{\epsilon}}_{\sigma}(\hat{\mathbf{k}}) \times \hat{\bm{\epsilon}}_{\sigma'}^*(\hat{\mathbf{k}}')\, ,  \;     \mathrm{from} \;  \, \eqref{s100}     . \label{k40d}
\end{align}
\end{subequations}
There are several features of \eqref{k30}-\eqref{k50}, and \eqref{k40a}-\eqref{k40d} which are worth highlighting.
\begin{enumerate}
  \item It is well known that the helicity $H$, the chirality $C$, and the spin angular momentum $\mathbf{S}$ are diagonal with respect to the helicity polarization basis. In fact, from \eqref{a130}, \eqref{b60}, and \eqref{s120} it follows that
\begin{subequations}
\begin{align}
H = & \;  \frac{2 \, \epsilon_0}{c}   \int \di^3 \mathbf{k} \, \frac{1}{\left| \mathbf{k} \right|} \Bigl[ \left| a_{-}(\mathbf{k}) \right|^2 - \left| a_{+}(\mathbf{k}) \right|^2 \Bigr] \, , \label{k60a} \\[6pt]
C  = & \; {2 \, \epsilon_0}  \int \di^3 \mathbf{k} \, \left| \mathbf{k} \right| \,  \Bigl[ \left| a_{-}(\mathbf{k}) \right|^2 - \left| a_{+}(\mathbf{k}) \right|^2 \Bigr]   \, , \label{k60b} \\[6pt]
\mathbf{S} = & \; \frac{2 \, \epsilon_0}{c}   \int \di^3 \mathbf{k} \,  \frac{\hat{\mathbf{k}}}{|\mathbf{k}|} \, \Bigl[ \left| a_{-}(\mathbf{k}) \right|^2 - \left| a_{+}(\mathbf{k}) \right|^2 \Bigr]   \,  . \label{k60d}
\end{align}
\end{subequations}
Conversely, the corresponding densities are not diagonal and cross-helicity terms do not vanish. However, from \eqref{k50} it follows that in the monochromatic limits the rapidly oscillating factors $\exp \left[ \pm i \left(\omega + \omega' \right) t \right]$ go to zero after averaging over a period of oscillation \cite{Aiello_2022}. The remaining  terms proportional to $\exp \left[ \pm i \left(\omega - \omega' \right) t \right]$ average to $1$, so that
\begin{align}
A_{\sigma \sigma '}(\mathbf{k},\mathbf{k}',t)
 \to & \;    a_\sigma(\mathbf{k})  a_{\sigma'}^*(\mathbf{k}') \left(\frac{\sigma' + \sigma}{2} \right) \nonumber \\[6pt]
 & =  \sigma \,\delta_{\sigma \sigma'} a_\sigma(\mathbf{k})  a_{\sigma}^*(\mathbf{k}') \, , \label{k70}
\end{align}
and the cross-helicity terms disappear.
  \item Equations \eqref{k40b} and \eqref{k40c} give another view on the non-uniqueness of the densities. Note, however, that the two expressions coincide for monochromatic light where $\left| \mathbf{k} \right| = \left| \mathbf{k}' \right| $.
  \item Equation \eqref{k40d} shows that the spin density possesses both a longitudinal and a transverse part, because from \eqref{eq280b} it follows that
\begin{align}
\mathbf{s}_{\sigma \sigma '}(\mathbf{k},\mathbf{k}') = & \; -i \, \sigma \, \hat{\bm{\epsilon}}_{\sigma}(\hat{\mathbf{k}}) \times \hat{\bm{\epsilon}}_{\sigma'}^*(\hat{\mathbf{k}}')  \nonumber \\[6pt]
= & \; \hat{\mathbf{k}}   \left[ \hat{\bm{\epsilon}}_{\sigma}(\hat{\mathbf{k}})  \cdot \hat{\bm{\epsilon}}_{\sigma'}^*(\hat{\mathbf{k}}') \right] \nonumber \\[6pt]
& - \hat{\bm{\epsilon}}_{\sigma}(\hat{\mathbf{k}}) \left[ \hat{\mathbf{k}} \cdot \hat{\bm{\epsilon}}_{\sigma'}^*(\hat{\mathbf{k}}') \right]  \, . \label{k80}
\end{align}
However, the spin $\mathbf{S}$ is purely longitudinal because spatial integration yields a delta function $\delta(\mathbf{k} - \mathbf{k}')$, so that the transverse part disappears, due to the transverse character of the electromagnetic field.
\end{enumerate}
 {
\section{Check of the consistency with previous results}\label{GaugeInv}
}

 {
Equation \eqref{S140} gives the optical spin $\mathbf{S}$  in terms of the observable electric and magnetic fields. Since the total optical angular momentum $\mathbf{J}$ is a gauge-invariant quantity, it follows that also the orbital optical angular momentum  $\mathbf{L} = \mathbf{J} - \mathbf{S}$ must be expressible in terms of the electric and magnetic fields only. This gauge-invariant split between orbital and spin optical angular momenta has been extensively discussed, for example, by Stewart \cite{Stewart_2005} and by Barnett, Cameron and coworkers \cite{Barnett_2010,Cameron_2012bis,Cameron_2013,Barnett_2016}. However, while Stewart obtained expressions for $\mathbf{L}$ and $\mathbf{S}$ in terms of  $\mathbf{E}$ and $\mathbf{B}$ only (thus obtaining manifestly gauge-invariant quantities), Barnett, Cameron and coworkers wrote $\mathbf{L}$ and $\mathbf{S}$ in terms of both the observable fields $\mathbf{E}$ and $\mathbf{B}$ and the transverse vector potentials $\mathbf{A} = \mathbf{A}^\perp(\mathbf{r},t)$ and $\mathbf{C} = \mathbf{C}^\perp(\mathbf{r},t)$. Our approach in the present paper is somewhere in between Barnett's and Stewart's, because we start from the ``Barnett-like'' spin density \eqref{e20c} (see Eq. (25) in \cite{Barnett_2016}), and from this we derive $\mathbf{S}$ in terms of  $\mathbf{E}$ and $\mathbf{B}$ only, in a ``Stewart-like'' way (see Eq. (15) in \cite{Stewart_2005}). Therefore, to complete the present paper, we must demonstrate the equivalence between our results about the optical spin $\mathbf{S}$, and the ones from Stewart and Barnett, Cameron and coworkers   \cite{referee}.
}

 {
\subsection{Stewart's approach}\label{Stewart}
}

 {
To begin with, we write the linear momentum density $\mathbf{p}(\mathbf{r},t)$ of an optical field, in the electric-magnetic democratic form:
\begin{align}
\mathbf{p}(\mathbf{r},t) = \frac{\varepsilon_0}{2}  \left[ \mathbf{E}(\mathbf{r},t) \times \mathbf{B}(\mathbf{r},t) - \mathbf{B}(\mathbf{r},t) \times \mathbf{E}(\mathbf{r},t) \right]. \label{p50}
\end{align}
In \cite{Aiello_2022}, using Helmholtz's decomposition theorem, we have shown that $\mathbf{E}$ and $\mathbf{B}$ can be expressed in terms of each other, as
\begin{subequations}\label{p60}
\begin{align}
\mathbf{E} (\mathbf{r},t) &  = \; -\frac{1}{4 \pi} \, \bnabla \times \int \di^3 \mathbf{r}' \, \frac{ \partial \, \mathbf{B}(\mathbf{r}',t)}{\partial t}  \frac{1 }{\left|\mathbf{r} - \mathbf{r}'  \right|}, \label{p60a}  \\[6pt]
\mathbf{B} (\mathbf{r},t) & = \;  \frac{1}{4 \pi c^2} \,  \bnabla \times \int \di^3 \mathbf{r}' \, \frac{ \partial \, \mathbf{E}(\mathbf{r}',t)}{\partial t}  \frac{1 }{\left|\mathbf{r} - \mathbf{r}'  \right|}. \label{p60b}
\end{align}
\end{subequations}
Substituting \eqref{p60b} and  \eqref{p60a} in the first and second term of \eqref{p50}, respectively, we obtain after a straightforward calculation
\begin{align}\label{p70}
\mathbf{p}(\mathbf{r},t) = & \; \frac{\epsilon_0}{8 \pi  c^2} \left[\bm{\Pi}_E(\mathbf{r},t) + \bm{\Pi}_B(\mathbf{r},t) \right],
\end{align}
where we have defined
\begin{align}\label{p80}
\bm{\Pi}_E(\mathbf{r},t) \equiv & \; \mathbf{E}(\mathbf{r},t) \times \left[ \bnabla \times \mathbf{F}(\mathbf{r},t) \right] \nonumber \\[6pt]
= & \; \sum_{i=1}^3 E_i \bigl( \bnabla F_i \bigr) - \bigl(\mathbf{E} \cdot \bnabla \bigr)\mathbf{F},
\end{align}
with $\mathbf{F}  = \mathbf{F}(\mathbf{r},t) $, defined by
\begin{align}\label{p90}
\mathbf{F}(\mathbf{r},t) = \int \di^3 \mathbf{r}' \, \frac{ \partial \, \mathbf{E}(\mathbf{r}',t)}{\partial t}  \frac{1 }{\left|\mathbf{r} - \mathbf{r}'  \right|}.
\end{align}
As usual, $\bm{\Pi}_B$ is obtained replacing $\mathbf{E}$ with $c \mathbf{B}$ everywhere, in the expressions of $\bm{\Pi}_E$ and $\mathbf{F}$.
}

 {
Next, following Stewart \cite{Stewart_2005}, we calculate the angular momentum $\mathbf{J} = \mathbf{J}_E + \mathbf{J}_B$, where
\begin{align}\label{p100}
\mathbf{J}_E =  \frac{\epsilon_0}{8 \pi  c^2} \int \di^3 \mathbf{r}  \bigl[ \mathbf{r} \times \mathbf{\Pi}_E(\mathbf{r},t) \bigr],
\end{align}
and $\mathbf{J}_B$ is obtained replacing $\mathbf{E}$ with $c \mathbf{B}$ everywhere in the equation above.
Substituting \eqref{p80} into \eqref{p100}, and using the following vector identity,
\begin{align}\label{p110}
\mathbf{r} \times \bigl( \mathbf{E} \cdot \bnabla \bigr) \mathbf{F} = \mathbf{F} \times \mathbf{E} + \bigl( \mathbf{E} \cdot \bnabla \bigr) \bigl( \mathbf{r} \times \mathbf{F} \bigr),
\end{align}
we obtain $\mathbf{J}_E = \mathbf{J}_{OE} + \mathbf{J}_{SE}$, where
\begin{align}\label{p120}
\mathbf{J}_{OE} =  \frac{\epsilon_0}{8 \pi  c^2} \int \di^3 \mathbf{r}  \sum_{i=1}^3 E_i \bigl( \mathbf{r} \times \bnabla \bigr)F_i ,
\end{align}
and
\begin{align}\label{p130}
\mathbf{J}_{SE} = & \;   \frac{\epsilon_0}{8 \pi  c^2} \int \di^3 \mathbf{r} \, \mathbf{E} \times \mathbf{F}  \nonumber \\[6pt]
& -  \frac{\epsilon_0}{8 \pi  c^2} \int \di^3 \mathbf{r} \,    \bigl( \mathbf{E} \cdot \bnabla \bigr) \bigl( \mathbf{r} \times \mathbf{F} \bigr).
\end{align}
Substituting \eqref{p90} into the first term of \eqref{p130}, we straightforwardly obtain  $\mathbf{S}$ as given by \eqref{S140}. It is not difficult to see that the second term in \eqref{p130} is equal to zero. In fact, defining $\mathbf{G} = \mathbf{r} \times \mathbf{F}$, we can rewrite
\begin{align}\label{p140}
\int \di^3 \mathbf{r} \; &     \bigl( \mathbf{E} \cdot \bnabla \bigr) \bigl( \mathbf{r} \times \mathbf{F} \bigr)_j \nonumber \\[6pt]
 = & \; \underbrace{\int \di x_2 \, \di x_3 \Bigl[ E_1 G_j \Bigr]_{x_1 = -\infty}^{x_1 = \infty} + \, \text{cyclic terms}}_{  \text{surface terms} \, = \, 0} \nonumber \\[6pt]
& -  \int \di^3 \mathbf{r} \,   \underbrace{ \bigl( \bnabla \cdot \mathbf{E} \bigr)}_{= \; 0 } G_j .
\end{align}
We have thus demonstrated that Stewart's optical spin $\mathbf{J}_{SE}$ coincides with  $\mathbf{S}$ given by \eqref{S140}.
}

The orbital optical angular momentum  can be calculated from \eqref{p120}. By setting $\mathbf{r} = \hat{e}_i \, x_i, \, (i=1,2,3)$, where here and hereafter summation over repeated indices is understood, we can calculate $\mathbf{J}_{OE}$  noting that
\begin{align}\label{p150}
 \bigl( \mathbf{r} \times \bnabla \bigr) F_i  = & \; \hat{e}_j \,  \varepsilon_{jkl} \, x_k \, \frac{\partial F_i}{\partial x_l}   \nonumber \\[6pt]
 =  & \; \hat{e}_j \,  \varepsilon_{jkl} \, x_k \int \di^3 \mathbf{r}' \, \frac{ \partial \, E_i(\mathbf{r}',t)}{\partial t} \underbrace{\frac{\partial}{\partial x_l}  \frac{1 }{\left|\mathbf{r} - \mathbf{r}'  \right|}}_{= \; - \frac{ x_l - x_l'}{ \abs{\mathbf{r} - \mathbf{r}'}^3}} \nonumber \\ 
 =  & \; \int \di^3 \mathbf{r}' \, \frac{ \partial \, E_i(\mathbf{r}',t)}{\partial t} \, \frac{\mathbf{r} \times \mathbf{r}'}{\abs{\mathbf{r} - \mathbf{r}'}^3},
\end{align}
where \eqref{p90} has been used and $\varepsilon_{jkl}$ is the Levi-Civita symbol \cite{PermutationSymbol}. Substituting \eqref{p150} into \eqref{p120}, we obtain
\begin{align}\label{p160}
\mathbf{J}_{OE} = & \;   \frac{\epsilon_0}{8 \pi  c^2} \int \di^3 \mathbf{r}  \nonumber \\[6pt]
&  \times \int \di^3 \mathbf{r}' \,\mathbf{E}(\mathbf{r},t) \cdot \frac{ \partial \, \mathbf{E}(\mathbf{r}',t)}{\partial t} \, \frac{\mathbf{r} \times \mathbf{r}'}{\abs{\mathbf{r} - \mathbf{r}'}^3},
\end{align}
which, \emph{mutatis mutandis}, reproduces Eq. (17) in \cite{Stewart_2005}.

\vspace{0.5truecm}

\subsection{The approach by Barnett, Cameron, and coworkers}\label{Barnett}

{
By definition, our expression \eqref{S140} for the optical spin  $\mathbf{S}$ coincides with the one given by the second Eq. (25) in \cite{Barnett_2016}, because we use the same spin density
\begin{align}\label{p170}
 \mathbf{s}(\mathbf{r},t)  =  \frac{1}{2} \Bigl[  \epsilon_0 \, \mathbf{E} \times \mathbf{A} + \mathbf{B} \times \mathbf{C} \Bigr]  .
\end{align}
So, the last thing left to prove is that  the expression for $\mathbf{J}_{O} = \mathbf{L}$ given by the first Eq. (25) in \cite{Barnett_2016}, that is
\begin{align}\label{p180}
\mathbf{L}  =  & \; \frac{1}{2} \int \di^3 \mathbf{r} \,\sum_{j=1}^3 \bigl[ \epsilon_0 E_j \left( \mathbf{r} \times \bnabla \right) A_j + B_j \left( \mathbf{r} \times \bnabla \right) C_j  \bigr] \nonumber \\[6pt]
\equiv & \; \mathbf{L}_{E} + \mathbf{L}_{B},
\end{align}
 coincides with $\mathbf{J}_{O} = \mathbf{J}_{OE} + \mathbf{J}_{OB}$ calculated from \eqref{p160}. For simplicity, we will verify explicitly only that $\mathbf{L}_{E} = \mathbf{J}_{OE}$. For this purpose, we use \eqref{e32b} and \eqref{e184} with $\mathbf{G} = \mathbf{B}$  to write
\begin{align}\label{p190}
\mathbf{A}^\perp(\mathbf{r},t)  =  & \; \bigl( \bnabla \times \bigr)^{-1} \mathbf{B}(\mathbf{r},t)  \nonumber \\[6pt]
=  & \; \int \frac{\di^3 \mathbf{r}'}{4 \pi} \, \frac{\bnabla'  \times \, \mathbf{B} (\mathbf{r}',t)}{\left|\mathbf{r} - \mathbf{r}'  \right|} \nonumber \\[6pt]
=  & \; \frac{1}{4 \pi c^2} \, \mathbf{F}(\mathbf{r},t),
\end{align}
where \eqref{S120} and \eqref{p90} have been used.
Substituting \eqref{p190} into the first term between square brackets in \eqref{p180}, we obtain
\begin{align}\label{p200}
\mathbf{L}_{E} =   \frac{\epsilon_0}{8 \pi c^2}  \int \di^3 \mathbf{r} \,\sum_{j=1}^3  E_j (\mathbf{r},t) \left( \mathbf{r} \times \bnabla \right) F_j(\mathbf{r},t)  ,
\end{align}
which coincides with \eqref{p120}. This completes the consistency check.
}

\section{Concluding remarks}\label{Conclusions}

In this paper I have presented some alternative expressions for the well-known  helicity $H$,  chirality $C$, and  spin angular momentum $\mathbf{S}$ of an optical field, based solely on  observable quantities of the electromagnetic field.  The  main results of this paper are summarized as follows.
\begin{enumerate}
  \item Equations \eqref{k20} and \eqref{k10a}-\eqref{k10c} give manifestly gauge-invariant expressions for the helicity, chirality, and spin \emph{densities} of the electromagnetic field, without using the auxiliary transverse vector potentials $\mathbf{A}$ and $\mathbf{C}$. The simple form of these equations makes clear the connection between $H$,  $C$, and  $\mathbf{S}$, and the electric-magnetic democracy is fully displayed.
  \item In the reciprocal space the helicity, chirality, and spin densities reveal their connection at a deeper level, as shown by \eqref{k30}-\eqref{k50} and \eqref{k40a}-\eqref{k40d}. These equations also highlight the special role played by monochromatic  fields.

 {
  \item The key to achieving these results is the inversion  of the curl operator via Helmholtz's decomposition  theorem \cite{Zangwill}. I have shown in Sec. \ref{Helmholtz} that while this inversion is ill-defined for arbitrary vector fields, it becomes perfectly well defined for solenoidal fields.
This makes it possible to obtain manifestly gauge-invariant electromagnetic observable quantities.

In this work, I use the inverse curl operator in two different manners. The first use is exemplified by \eqref{e32a} and \eqref{e32b}, where I write the transverse part of the electric and magnetic vector potentials $\mathbf{C}$ and $\mathbf{A}$, in terms of the observable electric and magnetic fields $\mathbf{E}$ and $\mathbf{B}$, respectively. The second usage is shown in \eqref{p60a}  and \eqref{p60b}, where the electric and magnetic field are written as a ``Helmholtz transform'' pair.

 It is perhaps surprising that such a useful and powerful technique is not so commonly used in angular momentum optics, with a few commendable exceptions \cite{Stewart_2003,Stewart_2004,Stewart_2005}.
}
\end{enumerate}
It would be desirable if the results presented here could stimulate further research into the physical interpretation of the (virtually infinite) conserved quantities of the electromagnetic field.

\acknowledgments
I acknowledge support from Deutsche Forschungsgemeinschaft Project No. 429529648-
TRR 306 QuCoLiMa (``Quantum Cooperativity of Light and Matter''). I am grateful to Iwo Bialynicki-Birula for some important tips on calculating an integral.

\appendix

\section{Notation}\label{appZero}

In this appendix we quickly review the notation used throughout this paper, following \cite{Aiello_2022}.
The electric and magnetic fields $\mathbf{E}$ and $\mathbf{B}$, respectively, are given by
\begin{subequations}
\begin{align}
\mathbf{E}(\mathbf{r},t) = & \! \int \! \frac{\di^3 \mathbf{k}}{(2 \pi)^{3/2}} \Bigl[ \bm{a}(\mathbf{k}) e^{i \left( \mathbf{k} \cdot \mathbf{r} - \omega t \right)} + \bm{a}^*(\mathbf{k}) e^{-i \left( \mathbf{k} \cdot \mathbf{r} - \omega t \right)}  \Bigr]  , \label{a10a} \\[6pt]
c \mathbf{B}(\mathbf{r},t) = & \! \int \! \frac{\di^3 \mathbf{k}}{(2 \pi)^{3/2}} \Bigl[ \bm{b}(\mathbf{k}) e^{i \left( \mathbf{k} \cdot \mathbf{r} - \omega t \right)} + \bm{b}^*(\mathbf{k}) e^{-i \left( \mathbf{k} \cdot \mathbf{r} - \omega t \right)}  \Bigr]     , \label{a10b}
\end{align}
\end{subequations}
where $\omega = c \left|\mathbf{k} \right|$, and
\begin{align}
\bm{b}(\mathbf{k}) = \hat{\mathbf{k}} \times \bm{a}(\mathbf{k})  . \label{a20}
\end{align}
The time-independent vector amplitude $\bm{a}(\mathbf{k})$ can be calculated as
\begin{align}
\bm{a}(\mathbf{k}) & = \frac{1}{2}   \int \frac{\di^3 r}{\left( 2 \pi \right)^{3/2}} \left[ \mathbf{E}(\mathbf{r},t) + \frac{i}{\omega} \frac{\partial}{\partial t} \, \mathbf{E}(\mathbf{r},t) \right] \nonumber \\[6pt]
& \times \exp \left( - i \mathbf{k} \cdot \mathbf{r} + i \omega t \right) \, . \label{zero10}
\end{align}

We use a Cartesian coordinate system with the three perpendicular axes parallel to the unit vectors $\hat{\bm{e}}_1(\hat{\mathbf{k}}), \hat{\bm{e}}_2(\hat{\mathbf{k}})$ and $\hat{\bm{e}}_3(\hat{\mathbf{k}}) = \hat{\mathbf{k}}$, such that
\begin{align}
\hat{\bm{e}}_a(\hat{\mathbf{k}}) \cdot \hat{\bm{e}}_b(\hat{\mathbf{k}}) = \delta_{ab}, \; \mathrm{and} \;
\hat{\bm{e}}_a(\hat{\mathbf{k}}) \times \hat{\bm{e}}_b(\hat{\mathbf{k}}) = \varepsilon_{abc} \, \hat{\bm{e}}_c(\hat{\mathbf{k}}) , \label{eq130}
\end{align}
where $\varepsilon_{abc}$ denotes the Levi-Civita symbol with $a,b,c \in \{ 1,2,3 \}$, and summation over repeated indices (Einstein's summation convention), is understood.
Using the transverse basis $\{ \hat{\bm{e}}_1(\hat{\mathbf{k}}),\hat{\bm{e}}_2(\hat{\mathbf{k}})\}$, we can  build the so-called helicity (or, circular) polarization basis \cite{Berry_414},  $\{ \hat{\bm{\epsilon}}_+(\hat{\mathbf{k}}) , \hat{\bm{\epsilon}}_-(\hat{\mathbf{k}})\}$, defined by
\begin{align}
\hat{\bm{\epsilon}}_\sigma(\hat{\mathbf{k}})= \frac{\hat{\bm{e}}_1(\hat{\mathbf{k}}) - i  \, \sigma \, \hat{\bm{e}}_2(\hat{\mathbf{k}})}{\sqrt{2}}, \qquad (\sigma = \pm 1). \label{eq156}
\end{align}
Note that according to this definition, $\hat{\bm{\epsilon}}_+(\hat{\mathbf{k}})$ represents right-hand circular polarization, and $\hat{\bm{\epsilon}}_-(\hat{\mathbf{k}})$ represents left-hand circular polarization.
These two orthogonal unit complex vectors  have the following properties:
\begin{subequations}
\begin{align}
\hat{\mathbf{k}} \cdot \hat{\bm{\epsilon}}_{\sigma}(\hat{\mathbf{k}}) = & \; 0 , \label{eq280a} \\[6pt]
 \hat{\mathbf{k}} \times \hat{\bm{\epsilon}}_{\sigma}(\hat{\mathbf{k}}) = & \;  i \, \sigma \, \hat{\bm{\epsilon}}_{\sigma}(\hat{\mathbf{k}}), \label{eq280b} \\[6pt]
\hat{\bm{\epsilon}}_\sigma^*(\hat{\mathbf{k}}) \cdot \hat{\bm{\epsilon}}_{\sigma'}(\hat{\mathbf{k}}) = & \;  \delta_{\sigma \sigma'} , \label{eq280c} \\[6pt]
\hat{\bm{\epsilon}}_{\sigma}^*(-\hat{\mathbf{k}}) = & \;  \hat{\bm{\epsilon}}_{\sigma}(\hat{\mathbf{k}}),
 \label{eq280d} \\[6pt]
\hat{\bm{\epsilon}}_\sigma(\hat{\mathbf{k}}) \times \hat{\bm{\epsilon}}_{\sigma'}^*(\hat{\mathbf{k}}) = & \;  i \, \hat{\mathbf{k}} \, \sigma' \, \delta_{\sigma \sigma'}\, , \label{eq280e}
\end{align}
\end{subequations}
where $\sigma, \sigma' = \pm 1$. We can write $\bm{a}(\mathbf{k}) $ and $\bm{b}(\mathbf{k}) $, in the helicity  basis as
\begin{subequations}
\begin{align}
\bm{a}(\mathbf{k})  & = \sum_{\sigma = \pm 1} a_\sigma(\mathbf{k}) \, \hat{\bm{\epsilon}}_\sigma(\hat{\mathbf{k}}) \, , \label{eq142a} \\[4pt]
\bm{b}(\mathbf{k}) & =  \sum_{\sigma = \pm 1} b_\sigma(\mathbf{k}) \, \hat{\bm{\epsilon}}_\sigma(\hat{\mathbf{k}}) \, ,  \label{eq142b}
\end{align}
\end{subequations}
where the components
\begin{subequations}
\begin{align}
a_\sigma(\mathbf{k})  & = \hat{\bm{\epsilon}}_\sigma^*(\hat{\mathbf{k}}) \cdot \bm{a}(\mathbf{k}), \label{eq143a} \\[4pt]
b_\sigma(\mathbf{k}) & =  \hat{\bm{\epsilon}}_\sigma^*(\hat{\mathbf{k}}) \cdot \bm{b}(\mathbf{k}) \label{eq143b}
\end{align}
\end{subequations}
satisfy the equation
\begin{align}
b_\sigma(\mathbf{k}) =  i \, \sigma \, a_\sigma(\mathbf{k}), \qquad (\sigma = \pm 1). \label{eq144}
\end{align}

\section{Helicity, chirality, and spin in reciprocal space}\label{appa}

In this appendix we calculate the expressions of $H$, $C$, and $\mathbf{S}$ in  reciprocal space, in terms of the Fourier transform of the electric and magnetic fields. In what follows, some attention should be paid to the fact that we use the same times symbol ``$\times$'', both to denote ordinary multiplication, as in $ 2 \times 3 = 6$, and the vector product between two vectors.

\subsection{Helicity in reciprocal space}\label{appaH}

It is convenient to rewrite \eqref{H20}  as $H = H_E + H_B$, where
\begin{align}
 H_E = & \;  \frac{\epsilon_0}{8 \pi c} \,  \int \di^3 \mathbf{r} \int \di^3 \mathbf{r}'  \, \frac{\mathbf{E}(\mathbf{r},t) \cdot \Bigl[ \bnabla'  \times  \mathbf{E}(\mathbf{r}',t) \Bigr] }{\left|\mathbf{r} - \mathbf{r}' \right|} \, , \label{a30}
\end{align}
and $H_B$ is obtained from \eqref{a30} replacing $\mathbf{E}$ with $c \mathbf{B}$, everywhere.
Next, substituting \eqref{a10a} into \eqref{a30},  we obtain
\begin{widetext}
\begin{align}
 H_E = & \;  \frac{\epsilon_0}{8 \pi c} \,  \int \di^3 \mathbf{r} \int \di^3 \mathbf{r}'  \, \frac{1}{\left|\mathbf{r} - \mathbf{r}' \right|} \int \frac{\di^3 \mathbf{k}}{(2 \pi)^{3/2}} \int \frac{\di^3 \mathbf{k}'}{(2 \pi)^{3/2}} \nonumber \\[6pt]
& \times    \Bigl[ \bm{a}(\mathbf{k}) e^{i \left( \mathbf{k} \cdot \mathbf{r} - \omega t \right)} + \bm{a}^*(\mathbf{k}) e^{-i \left( \mathbf{k} \cdot \mathbf{r} - \omega t \right)}  \Bigr]
\cdot \left\{ i \, \mathbf{k}'  \times
 \Bigl[  \bm{a}(\mathbf{k}') e^{i \left( \mathbf{k}' \cdot \mathbf{r}' - \omega' t \right)} - \bm{a}^*(\mathbf{k}') e^{-i \left( \mathbf{k}' \cdot \mathbf{r}' - \omega' t \right)} \Bigr] \right\}  \, , \label{a40}
\end{align}
where  \eqref{e80} has been used. Calculating the scalar product in the second and third line  above, we obtain after a little calculation
\begin{align}
 H_E = & \;   \frac{\epsilon_0}{8 \pi c} \,  \int \frac{\di^3 \mathbf{r}}{(2 \pi)^3} \int \di^3 \mathbf{r}'  \, \frac{1}{\left|\mathbf{r} - \mathbf{r}' \right|} \int \di^3 \mathbf{k} \int \di^3 \mathbf{k}' \, \left| \mathbf{k}'\right| \nonumber \\[6pt]
&  \times  \Bigl\{ \alpha\left( \mathbf{k} , \mathbf{k}'\right) \exp \left[i \left( \mathbf{r} \cdot \mathbf{k} + \mathbf{r}' \cdot\mathbf{k}'\right) - i \left( \omega + \omega' \right)t \right]  -\beta\left( \mathbf{k} , \mathbf{k}'\right) \exp \left[i \left( \mathbf{r} \cdot \mathbf{k} - \mathbf{r}' \cdot\mathbf{k}'\right)  - i \left( \omega - \omega' \right)t \right] \Bigr\} +  \mathrm{c.c.}
 \, , \label{a50}
\end{align}
where we have defined
\begin{subequations}
\begin{align}
\alpha\left( \mathbf{k} , \mathbf{k}'\right) = & \;  i \, \bm{a}(\mathbf{k}) \cdot \left[ \hat{\mathbf{k}}' \times     \bm{a}(\mathbf{k}') \right],  \label{a60a} \\[6pt]
\beta\left( \mathbf{k} , \mathbf{k}'\right) = & \; i \, \bm{a}(\mathbf{k}) \cdot \left[ \hat{\mathbf{k}}' \times     \bm{a}^*(\mathbf{k}') \right] \label{a60b} \, ,
\end{align}
\end{subequations}
with $\omega' = c \left|\mathbf{k}' \right| $,  and $\mathrm{c.c.}$ stands for c\emph{omplex} c\emph{onjugate}. It is not difficult to show that we can rewrite $H_E$ as
\begin{align}
 H_E = & \;   \frac{\epsilon_0}{8 \pi c} \,  \int \frac{\di^3 \mathbf{r}}{(2 \pi)^3}     \int \di^3 \mathbf{k} \, \int \di^3 \mathbf{k}' \, \bcancel{\left| \mathbf{k}'\right|}  \, e^{i   \mathbf{r} \cdot\mathbf{k} }  \nonumber \\[6pt]
&  \times  \Biggl( \Bigl\{  \alpha\left( \mathbf{k} , -\mathbf{k}'\right) \exp \left[- i \left( \omega + \omega' \right)t \right] -  \beta\left( \mathbf{k} , \mathbf{k}'\right) \exp \left[  - i \left( \omega - \omega' \right)t \right] \Bigr\}  \underbrace{\int \di^3 \mathbf{r}'  \, \frac{\displaystyle e^{-i   \mathbf{r}' \cdot\mathbf{k}' }}{\left|\mathbf{r} - \mathbf{r}' \right|}}_{\displaystyle = \; e^{-i   \mathbf{r} \cdot \mathbf{k}' } {4 \pi}/{\left|\mathbf{k}' \right|^{\bcancel{2}}}} \Biggr) +  \mathrm{c.c.} \nonumber \\[6pt]
= & \; \frac{\epsilon_0}{2 c} \, \int \frac{\di^3 \mathbf{r}}{(2 \pi)^3}     \int \di^3 \mathbf{k} \, \int \di^3 \mathbf{k}' \, \frac{ e^{i   \mathbf{r} \cdot \left(\mathbf{k} - \mathbf{k}' \right)}}{\left| \mathbf{k}'\right|}  \left[  \alpha \left( \mathbf{k} , -\mathbf{k}'\right) e^{- i \left( \omega + \omega' \right)t } -  \beta\left( \mathbf{k} , \mathbf{k}'\right) e^{- i \left( \omega - \omega' \right)t } \right]  +  \mathrm{c.c.}
 \, , \label{a70}
\end{align}
where we have made the change of variables $\mathbf{k}' \to -\mathbf{k}'$ in the part of the integrand proportional to $\alpha\left( \mathbf{k},\mathbf{k}'\right)$.
Using the helicity basis \eqref{eq156}, we can rewrite
\begin{align}\label{a80a}
 \alpha\left( \mathbf{k} , -\mathbf{k}'\right) =  - \sum_{\sigma, \sigma' = \pm 1} \sigma' a_\sigma(\mathbf{k}) a_{\sigma'}(-\mathbf{k}') \left[ \hat{\bm{\epsilon}}_{\sigma}(\hat{\mathbf{k}}) \cdot \hat{\bm{\epsilon}}_{\sigma'}^*(\hat{\mathbf{k}}') \right]  \, ,
\end{align}
and
\begin{align}
 \beta\left( \mathbf{k} , \mathbf{k}'\right) =   \sum_{\sigma, \sigma' = \pm 1} \sigma' a_\sigma(\mathbf{k}) a_{\sigma'}^*(\mathbf{k}') \left[\hat{\bm{\epsilon}}_{\sigma}(\hat{\mathbf{k}}) \cdot \hat{\bm{\epsilon}}_{\sigma'}^*(\hat{\mathbf{k}}') \right]  \label{a80b} \, ,
\end{align}
where \eqref{eq280c}-\eqref{eq280e} and \eqref{eq142a}, have been used.

Substituting \eqref{a80a} and \eqref{a80b} into \eqref{a70}, we eventually obtain
\begin{align}\label{a90}
 H_E = & \;    -\frac{\epsilon_0}{2 c} \sum_{\sigma, \sigma' = \pm 1} \int \frac{\di^3 \mathbf{r}}{(2 \pi)^3}     \int \di^3 \mathbf{k} \, \int \di^3 \mathbf{k}'\, \frac{ e^{i   \mathbf{r} \cdot \left(\mathbf{k} - \mathbf{k}' \right)}}{\left| \mathbf{k}'\right|}  \,  \hat{\bm{\epsilon}}_{\sigma}(\hat{\mathbf{k}}) \cdot \hat{\bm{\epsilon}}_{\sigma'}^*(\hat{\mathbf{k}}')  \nonumber \\[6pt]
&  \times \sigma'  a_\sigma(\mathbf{k})\left[  a_{\sigma'}(-\mathbf{k}')  e^{- i \left( \omega + \omega' \right)t } + a_{\sigma'}^*(\mathbf{k}') e^{- i \left( \omega - \omega' \right)t } \right]  +  \mathrm{c.c.} \,
 .
\end{align}
To calculate $H_B$ we simply take the expression above for $H_E$ and we replace $a_\sigma(\mathbf{k})$ with $b_\sigma(\mathbf{k})= i \sigma a_\sigma(\mathbf{k})$, according to \eqref{eq144}. This implies that
\begin{subequations}
\begin{align}
\sigma' a_\sigma(\mathbf{k}) a_{\sigma'}(-\mathbf{k}') &   \to    \sigma' b_\sigma(\mathbf{k}) b_{\sigma'}(-\mathbf{k}' ) = -\sigma a_\sigma(\mathbf{k}) a_{\sigma'}(-\mathbf{k}')  \, ,  \label{a100a} \\[6pt]
\sigma' a_\sigma(\mathbf{k}) a_{\sigma'}^*(\mathbf{k}') & \to   \sigma' b_\sigma(\mathbf{k}) b_{\sigma'}^*(\mathbf{k}') = \sigma a_\sigma(\mathbf{k}) a_{\sigma'}^*(\mathbf{k}') \label{a100b} \, .
\end{align}
\end{subequations}
From this result and \eqref{a90}, it follows that
\begin{align}
 H      = & \; H_E + H_B   \nonumber \\[6pt]
= & \;  -\frac{\epsilon_0}{c} \sum_{\sigma, \sigma' = \pm 1} \int \frac{\di^3 \mathbf{r}}{(2 \pi)^3}     \int \di^3 \mathbf{k} \, \int \di^3 \mathbf{k}' \, \frac{ e^{i   \mathbf{r} \cdot \left(\mathbf{k} - \mathbf{k}' \right)}}{\left| \mathbf{k}'\right|}  \,    \hat{\bm{\epsilon}}_{\sigma}(\hat{\mathbf{k}}) \cdot \hat{\bm{\epsilon}}_{\sigma'}^*(\hat{\mathbf{k}}')  \nonumber \\[6pt]
&  \times a_\sigma(\mathbf{k}) \left\{  a_{\sigma'}(-\mathbf{k}')\left(\frac{\sigma' - \sigma }{2}\right)   \exp\left[{- i \left( \omega + \omega' \right)t }\right] + \, a_{\sigma'}^*(\mathbf{k}') \left(\frac{\sigma' + \sigma}{2} \right) \exp\left[{- i \left( \omega - \omega' \right)t }\right] \right\}  +  \mathrm{c.c.} \,  . \label{a110}
\end{align}
\end{widetext}
Integration of this expression with respect to $\di^3 \mathbf{r}$ yields $(2 \pi)^3$ times the delta function $\delta \left( \mathbf{k}' - \mathbf{k} \right)$, so that
\begin{align}
\delta \left( \mathbf{k}' - \mathbf{k} \right) \hat{\bm{\epsilon}}_{\sigma}(\hat{\mathbf{k}}) \cdot \hat{\bm{\epsilon}}_{\sigma'}^*(\hat{\mathbf{k}}') = & \; \delta \left( \mathbf{k}' - \mathbf{k} \right) \delta_{\sigma \sigma'}\, .\label{a120}
\end{align}
Using this result, we directly obtain
\begin{align}
H = & \;  -\frac{\epsilon_0}{c} \sum_{\sigma = \pm 1}  \sigma   \int \frac{\di^3\mathbf{k}}{\left| \mathbf{k}\right|}  \,   \left| a_\sigma(\mathbf{k}) \right|^2 +  \mathrm{c.c.} \nonumber \\[6pt]
= & \;  \frac{2 \, \epsilon_0 }{c}   \int \frac{\di^3 \mathbf{k}}{ \left| \mathbf{k} \right| }  \,   \Bigl[ \left| a_-(\mathbf{k}) \right|^2 - \left| a_+(\mathbf{k}) \right|^2 \Bigr] \,
. \label{a130}
\end{align}

We show now that this result is in agreement with the expressions for $H$ given in the literature in terms of photon-number operators (see, e.g., Eq. (2.8) in \cite{Cameron_2012}). First, we use our notation to rewrite Eq. (10.4-39) in \cite{MandelBook}, which  gives the electric-field quantum operator in a finite quantization volume $L^3$, as
\begin{align}
\hat{\mathbf{E}}(\mathbf{r},t) = & \;   \sum_{\mathbf{k}, \sigma}  \sqrt{\frac{\hbar \, \omega}{2 \epsilon_0 L^3} }  \, \Bigl[ i \, \hat{a}_{\mathbf{k} \sigma} \, \bm{\epsilon}_{\mathbf{k} \sigma} e^{i \left( \mathbf{k} \cdot \mathbf{r} - \omega t \right)}  +  \mathrm{h.c.} \Bigr]  \,
, \label{a140}
\end{align}
where h.c. stands for the Hermitian conjugate of the preceding term. To compare this expression with $\mathbf{E}(\mathbf{r},t)$ given by \eqref{a10a}, we must convert the integral in  \eqref{a10a}, to a sum according to the general rule (see, e.g., Eq. (10.8-4) in \cite{MandelBook}),
\begin{align}
\sum_{\mathbf{k}} \to \left( \frac{L}{2 \pi} \right)^{3}  \int \di^3 \mathbf{k}   \, , \label{a150}
\end{align}
so that \eqref{a10a} becomes
\begin{multline}
\mathbf{E}(\mathbf{r},t) \\[6pt]
= \frac{\left(2 \pi \right)^{3/2}}{L^3} \,\sum_{\mathbf{k}, \sigma}  \left[ a_\sigma(\mathbf{k}) \hat{\epsilon}_\sigma(\hat{\mathbf{k}}) e^{i \left( \mathbf{k} \cdot \mathbf{r} - \omega t \right)} + \mathrm{c.c.} \right] \, , \label{a10aBIS}
\end{multline}
where \eqref{eq142a} has been used.
From the comparison between \eqref{a140} and \eqref{a10aBIS}, it follows that
\begin{align}
\mathbf{E}(\mathbf{r},t) \to \hat{\mathbf{E}}(\mathbf{r},t), \; \, \mathrm{if} \; \, a_\sigma(\mathbf{k})  \to  \left[ \frac{\hbar \, \omega L^3}{2(2 \pi)^3 \epsilon_0} \right]^{1/2}   \hat{a}_{\mathbf{k} \sigma}  \, . \label{a142}
\end{align}
This implies that we can define the  classical quantities $n_{\mathbf{k}\sigma} $ corresponding to the quantum photon-number operators $\hat{n}_{\mathbf{k} \sigma}$, as
\begin{align}
n_{\mathbf{k}\sigma} \equiv  \frac{2(2 \pi)^3 \epsilon_0}{\hbar \omega L^3}   \left| a_\sigma (\mathbf{k}) \right|^2 \to  \hat{a}_{\mathbf{k} \sigma}^\dagger \hat{a}_{\mathbf{k} \sigma} \equiv \hat{n}_{\mathbf{k} \sigma} \, , \label{a170}
\end{align}
So, if we rename $n_{\mathbf{k}-} $ and $n_{\mathbf{k}+} $, as $n_{\mathbf{k}L} $ and $n_{\mathbf{k}R} $, respectively, where the subscripts $L$ and $R$ label left- and right-handed circular polarization,
 we can rewrite \eqref{a130} with the help of \eqref{a150}, as
\begin{align}
H = \sum_{\mathbf{k}} \hbar  \left( n_{\mathbf{k}L} - n_{\mathbf{k}R} \right)
, \label{a180}
\end{align}
in agreement with \cite{Cameron_2012}.

\subsection{Chirality in reciprocal space}\label{appb}

We have two distinct expressions for $C$, given by \eqref{C5} and \eqref{C70}. Here we rewrite both equations in the reciprocal space.

\subsubsection{Chirality in reciprocal space from \eqref{C5}.}\label{appb1}

The expression of
\begin{align}
C_E  = & \; \frac{\epsilon_0}{2}  \int \di^3 \mathbf{r} \, \mathbf{E} \cdot \left( \bnabla  \times  \mathbf{E} \right) \, , \label{b10}
\end{align}
in the reciprocal space, can be directly calculated substituting \eqref{a10a} into \eqref{b10}, as follows:
\begin{align}
C_E = & \;  \frac{\epsilon_0}{2} \,  \int \di^3 \mathbf{r}  \int \frac{\di^3 \mathbf{k}}{(2 \pi)^{3/2}} \int \frac{\di^3 \mathbf{k}'}{(2 \pi)^{3/2}} \nonumber \\[6pt]
 & \times  \Bigl[ \bm{a}(\mathbf{k}) e^{i \left( \mathbf{k} \cdot \mathbf{r} - \omega t \right)} + \bm{a}^*(\mathbf{k}) e^{-i \left( \mathbf{k} \cdot \mathbf{r} - \omega t \right)}  \Bigr] \nonumber \\[6pt]
& \phantom{\times  } \cdot \biggl\{ i \, \mathbf{k}'  \times
 \Bigl[  \bm{a}(\mathbf{k}') e^{i \left( \mathbf{k}' \cdot \mathbf{r} - \omega' t \right)} \nonumber \\[6pt]
 & \phantom{\times  \cdot \biggl\{  } - \bm{a}^*(\mathbf{k}') e^{-i \left( \mathbf{k}' \cdot \mathbf{r} - \omega' t \right)} \Bigr] \biggr\}  \, , \label{b20}
\end{align}
where \eqref{e80} has been used.  The evaluation of the scalar product in  \eqref{b20} above gives
\begin{align}
C_E = & \;   \frac{\epsilon_0}{2} \,  \int \frac{\di^3 \mathbf{r}}{(2 \pi)^3}     \int \di^3 \mathbf{k} \, \int \di^3 \mathbf{k}' \, \left| \mathbf{k}'\right| \, e^{i \mathbf{r} \cdot \left( \mathbf{k} - \mathbf{k}'\right) } \nonumber \\[6pt]
&  \text{\textsf{x}}  \,  \Bigl\{ \alpha\left( \mathbf{k} , -\mathbf{k}'\right) \exp \left[ - i \left( \omega + \omega' \right)t \right] \nonumber \\[6pt]
& \phantom{ \times  \Biggl( \Bigl\{  }  -\beta\left( \mathbf{k} , \mathbf{k}'\right) \exp \left[  - i \left( \omega - \omega' \right)t \right] \Bigr\} +  \mathrm{c.c.}
 \, , \label{b30}
\end{align}
where \eqref{a60a} and \eqref{a60b} have been used. Next, substituting \eqref{a80a} and \eqref{a80b} into \eqref{b30}, we obtain
\begin{align}
 C_E = & \;    -\frac{\epsilon_0}{2} \sum_{\sigma, \sigma' = \pm 1} \int \frac{\di^3 \mathbf{r}}{(2 \pi)^3}     \int \di^3 \mathbf{k} \, \int \di^3 \mathbf{k}'  \left| \mathbf{k}'\right|  e^{ i \mathbf{r} \cdot \left( \mathbf{k} - \mathbf{k}'\right) }\nonumber \\[6pt]
 & \times \hat{\bm{\epsilon}}_{\sigma}(\hat{\mathbf{k}}) \cdot \hat{\bm{\epsilon}}_{\sigma'}^*(\hat{\mathbf{k}}') \, \sigma'  a_\sigma(\mathbf{k}) \nonumber \\[6pt]
&  \times\left[  a_{\sigma'}(-\mathbf{k}')  e^{- i \left( \omega + \omega' \right)t } + a_{\sigma'}^*(\mathbf{k}') e^{- i \left( \omega - \omega' \right)t } \right] \nonumber \\[6pt]
& +  \mathrm{c.c.} \,  . \label{b40}
\end{align}
To calculate $C_B$  we replace $a_\sigma(\mathbf{k})$ with $b_\sigma(\mathbf{k})= i \sigma a_\sigma(\mathbf{k})$, in \eqref{b40}. Using \eqref{a100a} and \eqref{a100b} we can eventually write
\begin{align}
C  = & \; C_E + C_B   \nonumber \\[6pt]
= & \;  -\epsilon_0 \sum_{\sigma, \sigma' = \pm 1} \int \frac{\di^3 \mathbf{r}}{(2 \pi)^3}     \int \di^3 \mathbf{k} \, \int \di^3 \mathbf{k}' \, \frac{e^{i   \mathbf{r} \cdot \left(\mathbf{k} - \mathbf{k}' \right)}}{\left| \mathbf{k}'\right|}\left| \mathbf{k}'\right|^2 \nonumber  \\[6pt]
& \times    \hat{\bm{\epsilon}}_{\sigma}(\hat{\mathbf{k}}) \cdot \hat{\bm{\epsilon}}_{\sigma'}^*(\hat{\mathbf{k}}')  \, a_\sigma(\mathbf{k}) \nonumber \\[6pt]
&  \times  \left\{  a_{\sigma'}(-\mathbf{k}')\left(\frac{\sigma' - \sigma }{2}\right)   \exp\left[{- i \left( \omega + \omega' \right)t }\right] \right. \nonumber \\[6pt]
&  \phantom{a_\sigma(\mathbf{k})xxx } \left. + \, a_{\sigma'}^*(\mathbf{k}') \left(\frac{\sigma' + \sigma}{2} \right) \exp\left[{- i \left( \omega - \omega' \right)t }\right] \right\} \nonumber \\[6pt]
& +  \mathrm{c.c.} \,  . \label{b50}
\end{align}
Performing the integration in real space with respect to $\di^3 \mathbf{r} $, we obtain
\begin{align}
C = & \;  -\epsilon_0 \sum_{\sigma = \pm 1}  \sigma   \int \di^3 \mathbf{k} \,  \left| \mathbf{k}\right|  \left| a_{\sigma}(\mathbf{k}) \right|^2  +  \mathrm{c.c.} \nonumber \\[6pt]
= & \;  2 \, \epsilon_0      \int \di^3 \mathbf{k} \,  \left| \mathbf{k}\right|  \Bigl[ \left| a_{-}(\mathbf{k}) \right|^2 - \left| a_{+}(\mathbf{k}) \right|^2  \Bigr] \, . \label{b60}
\end{align}
Finally, using \eqref{a150}-\eqref{a170} it is not difficult to show that we can write the chirality in a quantum-like language as
\begin{align}
C = \sum_{\mathbf{k}}  \hbar \, c  \left| \mathbf{k}\right|^2  \left( n_{ \mathbf{k}L} - n_{ \mathbf{k}R} \right) \, . \label{b70}
\end{align}

\subsubsection{Chirality in reciprocal space from \eqref{C70}.} \label{appb2}

In this case it is not necessary to make new calculations because by comparing \eqref{C70} with \eqref{a30}, we can see that
\begin{align}
C_E = \left. -\frac{1}{c} H_E \right|_{\mathbf{E}(\mathbf{r},t) \to \frac{\partial^2 \mathbf{E}(\mathbf{r},t)}{\partial t^2}} \, . \label{b80}
\end{align}
Then, we can use this relation and \eqref{a110} to write directly
\begin{widetext}
\begin{align}
C  = & \;  -\epsilon_0 \sum_{\sigma, \sigma' = \pm 1} \int \frac{\di^3 \mathbf{r}}{(2 \pi)^3}     \int \di^3 \mathbf{k} \, \int \di^3 \mathbf{k}'\, \frac{e^{i   \mathbf{r} \cdot \left(\mathbf{k} - \mathbf{k}' \right)}}{\left| \mathbf{k}'\right|}\left| \mathbf{k}\right|^2      \hat{\bm{\epsilon}}_{\sigma}(\hat{\mathbf{k}}) \cdot \hat{\bm{\epsilon}}_{\sigma'}^*(\hat{\mathbf{k}}')  \nonumber \\[6pt]
&  \times a_\sigma(\mathbf{k}) \left\{  a_{\sigma'}(-\mathbf{k}')\left(\frac{\sigma' - \sigma }{2}\right)   \exp\left[{- i \left( \omega + \omega' \right)t }\right]  + \, a_{\sigma'}^*(\mathbf{k}') \left(\frac{\sigma' + \sigma}{2} \right) \exp\left[{- i \left( \omega - \omega' \right)t }\right] \right\}  +  \mathrm{c.c.} \,
 . \label{b90}
\end{align}

\subsection{Spin in reciprocal space}\label{appc}

We calculate
\begin{align}
\mathbf{S} = & \;   \frac{\epsilon_0}{8 \pi \, c^2} \, \int \di^3 \mathbf{r} \int \di^3 \mathbf{r}'  \, \frac{\displaystyle \mathbf{E}(\mathbf{r},t)   \times   \frac{\partial \mathbf{E}(\mathbf{r}',t)}{\partial t }}{\left|\mathbf{r} - \mathbf{r}' \right|}  + \frac{\epsilon_0}{8 \pi \, c^2} \, \int \di^3 \mathbf{r} \int \di^3 \mathbf{r}'  \,
\frac{\displaystyle c^2 \, \mathbf{B}(\mathbf{r},t)  \times    \frac{\partial \mathbf{B}(\mathbf{r}',t)}{\partial t }
}{\left|\mathbf{r} - \mathbf{r}' \right|} \nonumber \\[6pt]
 \equiv & \; \mathbf{S}_E + \mathbf{S}_B\, , \label{s10}
\end{align}
in reciprocal space.
Substituting \eqref{a10a} into \eqref{s10}, we obtain
\begin{align}
\mathbf{S}_E = & \;  \frac{\epsilon_0}{8 \pi c^2} \,  \int \di^3 \mathbf{r} \int \di^3 \mathbf{r}'  \, \frac{1}{\left|\mathbf{r} - \mathbf{r}' \right|} \int \frac{\di^3 \mathbf{k}}{(2 \pi)^{3/2}} \int \frac{\di^3 \mathbf{k}'}{(2 \pi)^{3/2}} \left(-  i c \left| \mathbf{k}' \right| \right)  \nonumber \\[6pt]
&  \times  \Bigl[ \bm{a}(\mathbf{k}) e^{i \left( \mathbf{k} \cdot \mathbf{r} - \omega t \right)} + \bm{a}^*(\mathbf{k}) e^{-i \left( \mathbf{k} \cdot \mathbf{r} - \omega t \right)}  \Bigr]   \times
 \Bigl[  \bm{a}(\mathbf{k}') e^{i \left( \mathbf{k}' \cdot \mathbf{r}' - \omega' t \right)} - \bm{a}^*(\mathbf{k}') e^{-i \left( \mathbf{k}' \cdot \mathbf{r}' - \omega' t \right)} \Bigr]    \, . \label{s20}
\end{align}
Performing the vector products, we obtain
\begin{align}
\mathbf{S}_E = &    -\frac{\epsilon_0}{8 \pi c} \,  \int \frac{\di^3 \mathbf{r}}{( 2 \pi )^3} \int \di^3 \mathbf{r}'  \, \frac{1}{\left|\mathbf{r} - \mathbf{r}' \right|} \int \di^3 \mathbf{k} \int \di^3 \mathbf{k}' \, \left| \mathbf{k}'\right| \nonumber \\[6pt]
&  \times  \Bigl\{ \bm{\alpha} \left( \mathbf{k} , \mathbf{k}'\right) \exp \left[i \left( \mathbf{r} \cdot \mathbf{k} + \mathbf{r}' \cdot\mathbf{k}'\right) - i \left( \omega + \omega' \right)t \right]   -\bm{\beta}\left( \mathbf{k} , \mathbf{k}'\right) \exp \left[i \left( \mathbf{r} \cdot \mathbf{k} - \mathbf{r}' \cdot\mathbf{k}'\right)  - i \left( \omega - \omega' \right)t \right] \Bigr\}  +  \mathrm{c.c.}
 \, , \label{s30}
\end{align}
where we have defined
\begin{subequations}
\begin{align}
\bm{\alpha} \left( \mathbf{k} , \mathbf{k}'\right) = & \;  i \, \bm{a}(\mathbf{k})   \times    \bm{a}(\mathbf{k}') ,  \label{s40a} \\[6pt]
\bm{\beta} \left( \mathbf{k} , \mathbf{k}'\right) = & \; i \, \bm{a}(\mathbf{k})  \times   \bm{a}^*(\mathbf{k}') \label{s40b} \, .
\end{align}
\end{subequations}
A few more calculations give
\begin{align}
\mathbf{S}_E = & \;   -\frac{\epsilon_0}{8 \pi c} \,  \int \frac{\di^3 \mathbf{r}}{(2 \pi)^3}     \int \di^3 \mathbf{k} \, \int \di^3 \mathbf{k}' \, \bcancel{\left| \mathbf{k}'\right|}  \, e^{i   \mathbf{r} \cdot\mathbf{k} }  \nonumber \\[6pt]
&  \times  \Biggl( \Bigl\{  \bm{\alpha} \left( \mathbf{k} , -\mathbf{k}'\right) \exp \left[- i \left( \omega + \omega' \right)t \right] -  \bm{\beta} \left( \mathbf{k} , \mathbf{k}'\right) \exp \left[  - i \left( \omega - \omega' \right)t \right] \Bigr\}   \underbrace{\int \di^3 \mathbf{r}'  \, \frac{\displaystyle e^{-i   \mathbf{r}' \cdot \mathbf{k}' }}{\left|\mathbf{r} - \mathbf{r}' \right|}}_{\displaystyle = \; e^{-i   \mathbf{r} \cdot \mathbf{k}' } {4 \pi}/{\left|\mathbf{k}' \right|^{\bcancel{2}}}} \Biggr) +  \mathrm{c.c.} \nonumber \\[6pt]
= & \; -\frac{\epsilon_0}{2 c} \, \int \frac{\di^3 \mathbf{r}}{(2 \pi)^3}     \int \di^3 \mathbf{k} \, \int \di^3 \mathbf{k}' \, \frac{ e^{i   \mathbf{r} \cdot \left(\mathbf{k} - \mathbf{k}' \right)}}{\left| \mathbf{k}'\right|}  \left[  \bm{\alpha} \left( \mathbf{k} , -\mathbf{k}'\right) e^{- i \left( \omega + \omega' \right)t } -  \bm{\beta} \left( \mathbf{k} , \mathbf{k}'\right) e^{- i \left( \omega - \omega' \right)t } \right]  +  \mathrm{c.c.}
 \, , \label{s50}
\end{align}
where we made the change of variables $\mathbf{k}' \to -\mathbf{k}'$ in the part of the integrand proportional to $\bm{\alpha} \left( \mathbf{k},\mathbf{k}'\right)$.

In the helicity basis \eqref{eq156}, we can write
\begin{subequations}
\begin{align}
 \bm{\alpha} \left( \mathbf{k} , -\mathbf{k}'\right) = & \; \displaystyle i \sum_{\sigma, \sigma' = \pm 1} a_\sigma(\mathbf{k}) a_{\sigma'}(-\mathbf{k}') \left[\hat{\bm{\epsilon}}_{\sigma}(\hat{\mathbf{k}})  \times   \hat{\bm{\epsilon}}_{\sigma'}^*(\hat{\mathbf{k}}') \right]  \, ,  \label{s60a} \\[6pt]
 \bm{\beta} \left( \mathbf{k} , \mathbf{k}'\right) = & \; \displaystyle i \sum_{\sigma, \sigma' = \pm 1}  a_\sigma(\mathbf{k}) a_{\sigma'}^*(\mathbf{k}') \left[\hat{\bm{\epsilon}}_{\sigma}(\hat{\mathbf{k}})  \times   \hat{\bm{\epsilon}}_{\sigma'}^*(\hat{\mathbf{k}}') \right]  \label{s60b} \, ,
\end{align}
\end{subequations}
where \eqref{eq280e} and \eqref{eq142a} have been used.  Next, substituting \eqref{s60a} and \eqref{s60b} into \eqref{s50}, we obtain
\begin{align}
\mathbf{S}_E = & \;    -\frac{\epsilon_0}{2 c} \sum_{\sigma, \sigma' = \pm 1} \int \frac{\di^3 \mathbf{r}}{(2 \pi)^3}     \int \di^3 \mathbf{k} \, \int \di^3 \mathbf{k}' \,  \frac{ e^{i   \mathbf{r} \cdot \left(\mathbf{k} - \mathbf{k}' \right)}}{\left| \mathbf{k}'\right|} \,\left[ i \, \hat{\bm{\epsilon}}_{\sigma}(\hat{\mathbf{k}})  \times   \hat{\bm{\epsilon}}_{\sigma'}^*(\hat{\mathbf{k}}') \right] \nonumber \\[6pt]
&  \times   a_\sigma(\mathbf{k})\left[  a_{\sigma'}(-\mathbf{k}')  e^{- i \left( \omega + \omega' \right)t } - a_{\sigma'}^*(\mathbf{k}') e^{- i \left( \omega - \omega' \right)t } \right]  +  \mathrm{c.c.} \, . \label{s70}
\end{align}
$\mathbf{S}_B$ is obtained from $\mathbf{S}_E$ by replacing $a_\sigma(\mathbf{k})$ with $b_\sigma(\mathbf{k})= i \sigma a_\sigma(\mathbf{k})$, in \eqref{s70}. Eventually, we obtain
\begin{align}
\mathbf{S} = & \; \mathbf{S}_E + \mathbf{S}_B   \nonumber \\[6pt]
= & \;  -\frac{\epsilon_0}{c} \sum_{\sigma, \sigma' = \pm 1} \int \frac{\di^3 \mathbf{r}}{(2 \pi)^3}     \int \di^3 \mathbf{k} \, \int \di^3 \mathbf{k}' \, \frac{ e^{i   \mathbf{r} \cdot \left(\mathbf{k} - \mathbf{k}' \right)}}{\left| \mathbf{k}'\right|} \,    \left[ i \, \hat{\bm{\epsilon}}_{\sigma}(\hat{\mathbf{k}})  \times  \hat{\bm{\epsilon}}_{\sigma'}^*(\hat{\mathbf{k}}') \right]  \nonumber \\[6pt]
&  \times  a_\sigma(\mathbf{k}) \left\{  a_{\sigma'}(-\mathbf{k}')\left(\frac{1 - \sigma \sigma'}{2}\right)   \exp\left[{- i \left( \omega + \omega' \right)t }\right] \right. \nonumber \\[6pt]
&  \phantom{a_\sigma(\mathbf{k})xxx } \left. - \, a_{\sigma'}^*(\mathbf{k}') \left(\frac{1 + \sigma \sigma'}{2} \right) \exp\left[{- i \left( \omega - \omega' \right)t }\right] \right\}  +  \mathrm{c.c.} \,
 . \label{s80}
\end{align}
To put this expression in a form useful for the comparison with \eqref{a110}, \eqref{b50}, and \eqref{b90}, we note that
\begin{align}
- \sigma  \left(\frac{ \sigma' -\sigma}{2} \right) = \left(\frac{1 - \sigma \sigma'}{2} \right)  \quad \mathrm{and} \quad \sigma  \left(\frac{ \sigma' + \sigma}{2} \right) = \left(\frac{1 + \sigma \sigma'}{2} \right) \, . \label{s90}
\end{align}
Substituting \eqref{s90} into \eqref{s80}, we obtain
\begin{align}
\mathbf{S}    = & \; \mathbf{S}_E + \mathbf{S}_B   \nonumber \\[6pt]
 = & \;  \frac{\epsilon_0}{c} \sum_{\sigma, \sigma' = \pm 1} \int \frac{\di^3 \mathbf{r}}{(2 \pi)^3}     \int \di^3 \mathbf{k} \, \int \di^3 \mathbf{k}' \, \frac{ e^{i   \mathbf{r} \cdot \left(\mathbf{k} - \mathbf{k}' \right)}}{\left| \mathbf{k}'\right|} \,    \left[ i \, \sigma \, \hat{\bm{\epsilon}}_{\sigma}(\hat{\mathbf{k}})  \times  \hat{\bm{\epsilon}}_{\sigma'}^*(\hat{\mathbf{k}}') \right]  \nonumber \\[6pt]
&  \times a_\sigma(\mathbf{k}) \left\{  a_{\sigma'}(-\mathbf{k}') \left( \frac{ \sigma' - \sigma}{2} \right)   \exp\left[{- i \left( \omega + \omega' \right)t }\right] + \, a_{\sigma'}^*(\mathbf{k}') \left( \frac{ \sigma' + \sigma}{2} \right)  \exp\left[{- i \left( \omega - \omega' \right)t }\right] \right\}  +  \mathrm{c.c.} \,
, \label{s100}
\end{align}
where, according to \eqref{eq280b},
\begin{align}
i \, \sigma \, \hat{\bm{\epsilon}}_{\sigma}(\hat{\mathbf{k}})  \times  \hat{\bm{\epsilon}}_{\sigma'}^*(\hat{\mathbf{k}}')  =
\hat{\bm{\epsilon}}_{\sigma}(\hat{\mathbf{k}}) \left[ \hat{\mathbf{k}} \cdot \hat{\bm{\epsilon}}_{\sigma'}^*(\hat{\mathbf{k}}') \right] - \hat{\mathbf{k}}   \left[ \hat{\bm{\epsilon}}_{\sigma}(\hat{\mathbf{k}})  \cdot \hat{\bm{\epsilon}}_{\sigma'}^*(\hat{\mathbf{k}}') \right]
\, . \label{s110}
\end{align}

As a last step, we perform the integration  with respect to $\di^3 \mathbf{r} $ in \eqref{s100}, to obtain, after a little calculation,
\begin{align}
\mathbf{S}  = & \; -\frac{\epsilon_0}{c} \sum_{\sigma = \pm 1}  \sigma   \int \di^3 \mathbf{k} \,  \frac{\mathbf{k}}{\left| \mathbf{k}\right|^2}  \left| a_{\sigma}(\mathbf{k}) \right|^2  +  \mathrm{c.c.} \nonumber \\[6pt]
= & \;  \frac{2 \, \epsilon_0}{c}   \int \di^3 \mathbf{k} \,  \frac{\hat{\mathbf{k}}}{|\mathbf{k}|} \, \Bigl[ \left| a_{-}(\mathbf{k}) \right|^2 - \left| a_{+}(\mathbf{k}) \right|^2 \Bigr] \, . \label{s120}
\end{align}
Finally, using \eqref{a150}-\eqref{a170} it is not difficult to show that we can write the spin in a quantum-like language as
\begin{align}
\mathbf{S} = \sum_{\mathbf{k}}  \hbar \, \hat{ \mathbf{k}}  \left( n_{ \mathbf{k}L} - n_{ \mathbf{k}R} \right) \, . \label{s130}
\end{align}

\end{widetext}

\end{document}